\documentclass[a4paper, 11pt]{article}  
\usepackage{geometry, graphicx, blindtext, authblk, subcaption}     
\usepackage{amssymb, amsmath, algorithmic, algorithm}
\usepackage[table]{xcolor}
\usepackage[colorlinks=true, linkcolor=blue, citecolor = blue]{hyperref}
\usepackage{booktabs, color}
\usepackage[toc,title]{appendix}
\newcommand{\bx}{\mathbf{x}}
\newcommand{\Cor}{\mathbb{C}\text{orr}}
\newcommand{\Cov}{\mathbb{C}\text{ov}}
\newcommand{\Var}{\mathbb{V}}
\newcommand{\Exp}{\mathbb{E}}
\title{Sequential adaptive design for emulating costly computer codes}
\author{Hossein Mohammadi\thanks{Corresponding Author: h.mohammadi@exeter.ac.uk} }
\author{Peter Challenor}
\affil{Department of Mathematics and Statistics, Faculty of Environment, Science and Economy, University of Exeter}
\date{}							
\begin{document}
\maketitle
\begin{abstract}
Gaussian processes (GPs) are generally regarded as the gold standard surrogate model for emulating computationally expensive computer-based simulators. However, the problem of training GPs as accurately as possible with a minimum number of model evaluations remains challenging. We address this problem by suggesting a novel adaptive sampling criterion called VIGF (variance of improvement for global fit). The improvement function at any point is a measure of the deviation of the GP emulator from the nearest observed model output. At each iteration of the proposed algorithm, a new run is performed where VIGF is the largest. Then, the new sample is added to the design and the emulator is updated accordingly. A batch version of VIGF is also proposed which can save the user time when parallel computing is available. Additionally, VIGF is extended to the multi-fidelity case where the expensive high-fidelity model is predicted with the assistance of a lower fidelity simulator. This is performed via hierarchical kriging. The applicability of our method is assessed on a bunch of test functions and its performance is compared with several sequential sampling strategies. The results suggest that our method has a superior performance in predicting the benchmark functions in most cases. An implementation of VIGF is available in the dgpsi R package, which can be found on CRAN.
\end{abstract}
\section{Introduction}
\label{sec:introduction} 
This paper deals with adaptive design of experiments (DoE) for Gaussian process (GP) models \cite{GPML} in the context of emulating complex computer codes. This problem is of great importance to many applications where time-consuming simulators are employed to study a physical system; see e.g., \cite{parnianifard2019}. Due to computational burden of such models, we only have access to a limited number of simulation runs. As a result, conducting analyses such as uncertainty/sensitivity analyses \cite{saltelli2004} that require a huge number of model evaluations becomes impossible. One way around this computational complexity is to approximate the model by a cheap-to-evaluate surrogate model. A review of the most common surrogates can be found in \cite{gramacy2020}. Among them, GP emulators have become a standard in the field of computer experiments due to their flexibility and inherent estimation of prediction uncertainty \cite{sacks1989, GPML}. 

GPs are constructed based on a set of (carefully-designed) training runs called \emph{design points}. The locations of those points have a consequential effect on the predictive performance of GPs. It is desired to build an emulator as accurately as possible with a minimum number of calls to the costly simulator. This challenging task is an active area of research with important applications \cite{beck2016, dutta2020}. Typically, the sampling strategies proposed for tackling this problem can be divided into two main groups: 
\begin{itemize}
\item[(I)] \emph{One-shot DoE} where all the design locations are chosen at once, in advance of building the emulator. In this framework, all regions of the input space are treated as equally important, and points are spread across the space as uniformly as possible. Such \emph{space-filling} design can be generated using e.g., maximin or minimax \cite{johnson1990} criterion. A maximin-based design, for example, is attained by maximising the minimal distance between all candidate points. Latin hypercube sampling (LHS) \cite{mckay1979}, full factorial \cite{box1961}, orthogonal array \cite{owen1992}, and uniform designs \cite{fang2000} are examples of the one-shot DoE. Since the number of design points is decided beforehand in such DoE, there is the risk of under/oversampling \cite{garud2017} which is a critical  issue in sampling from computationally expensive simulators. A space-filling design can also be achieved using quasi-Monte Carlo sampling approaches such as Halton \cite{halton1960} or Sobol \cite{sobol1967} sequences. These methods, however, do not consider output information and differ from adaptive strategies explained below.
\item[(II)] \emph{Adaptive DoE} in which the choice of the next input value depends on the previously observed responses. Generally, adaptive sampling algorithms are more computationally demanding to implement compared to one-shot approaches. However, this added complexity becomes negligible when working with time-consuming numerical models, where each simulation run can take several days to complete. Furthermore, adaptive methods do not suffer from under/oversampling as we can stop them once the emulator reaches an adequate level of accuracy. This is a desirable property when emulating costly computer codes. Also, an adaptive design allows us to sample more points in ``interesting" regions, e.g., where the model response is highly nonlinear or changes rapidly. The focus of this paper is on the GP-based adaptive sampling. We refer the reader to recent papers \cite{fuhg2021, liu2018} for a comprehensive survey of the existing methods in this regard. 
\end{itemize}

 An adaptive sampling method generally needs to make a trade-off between two concepts: \emph{exploitation} and \emph{exploration}. The former aims to concentrate the search on identified interesting regions and sample more points there. The latter seeks poorly-represented parts of the domain to discover the interesting regions that have not yet been detected \cite{liu2018, garud2017}. For example, in \cite{aute2013, jiang2015, li2009, liu2015} the leave-one-out (LOO) cross-validation error accounts for exploitation, and a distance-based metric (often as a constrained optimisation) for exploration. The LOO error is obtained by removing a point from the training set, fitting a GP to the remaining data, and computing the difference between the actual response and its prediction. The magnitude of the LOO error at a design point is a measure of the sensitivity of the emulator to that point. In other words, the larger the LOO error is, the more important the design point becomes. Thus, the new sample should be taken near the point with the largest LOO error. In view of this, \cite{crombecq2011, xu2014} suggested the idea of partitioning the input space using the Voronoi tessellation \cite{voronoi1908}, and choosing the next design location in the cell with the largest LOO metric. 
 
 A common issue with the methods relying on LOO is that such error is only defined at the experimental design and needs to be estimated at untried locations. In \cite{mohammadi2022}, the expected squared LOO (ES-LOO) error is introduced to guide exploitation. The ES-LOO error is estimated at unobserved points by a second GP model, which imposes an extra cost. Another issue with the LOO-based adaptive approaches is that the new samples are often clustered around the existing design. To circumvent this problem, a modified version of \emph{expected improvement} is used in \cite{mohammadi2022}. The performance of the adaptive sampling algorithm based on the ES-LOO criterion is compared with our proposed method in Section \ref{sec:experiments}.

 Variance-based methods are a classic way of executing DoE sequentially. In this strategy, the GP predictive variance is considered as the prediction error, i.e., the difference between the predictive mean and truth. In GP modelling, a natural choice for a variance-based selection criterion is to maximise the mean squared error (MSE) \cite{jin2002, sacks1989}, often referred to as active learning Mackay (ALM) \cite{mackay1992, gramacy2020}, especially in the machine learning community. The MSE function is a measure of the distance between points relying on the canonical metric given by the GP covariance function \cite{marmin2018}. As a consequence, a sequential sampling based on maximising MSE can lead to a space-filling design as shown in Section \ref{sec:experiments}. It is notable that since the MSE criterion does not depend on the output function values (see Section \ref{sec:GP}), a sequential design based on MSE is actually non-adaptive.
 	
Another variance-based criterion is the integrated MSE (IMSE) \cite{picheny2010, sacks1989}, which represents the average accuracy of the emulator across the input space. In this approach, a new sampling point is selected where the IMSE is minimised, resulting in the maximum reduction of overall prediction uncertainty. Active learning Cohn (ALC) \cite{cohn1996, gramacy2020} is a sequential design method that selects the next sampling location by maximising the expected reduction in the GP predictive variance averaged over the design space \cite{gramacy2009}. It is important to note, however, that computing the integrals required for IMSE and ALC becomes increasingly expensive as dimensionality grows. A potential drawback of the variance-based methods is that they are likely to favour points on the boundaries of the domain where the prediction uncertainty is large. Such points may not be informative since models are not often precisely known on the boundaries, and error can be significant there \cite{gramacy2009}.

Entropy-based adaptive design is the last category of sampling strategies we review here. Entropy, first introduced by Shannon \cite{shannon1948}, can be used to quantify the uncertainty in the outcome of a random variable. Entropy is a concept in information theory, and a maximum entropy design maximises the information gained \cite{shewry1987}. In the GP paradigm, such design can be achieved by maximising the determinant of the covariance matrix of the sample points, and is analogous to a \emph{D-optimal} design \cite{chaloner1995}. The maximum entropy algorithm suits sequential sampling, and is equivalent to the maximum MSE approach if only one new location is selected at each iteration \cite{jin2002}. On a different note, entropy-based adaptive criteria have been employed in Bayesian optimisation \cite{henrandez-lobato2014}, and estimating contours of costly functions \cite{cole2022}. An alternative to the entropy criterion is \emph{mutual information} \cite{krause2008}; it measures the amount of information that one random process provides about another. In MICE (Mutual Information for Computer Experiments) \cite{beck2016}, mutual information is defined between the GP fitted to the training data, and the emulator constructed based on a set of candidate points. A point from the candidate set that provides the highest mutual information is then selected for the next evaluation. In the MICE algorithm, the search is performed over a discrete space and, hence, it may not be as efficient as methods defined on a continuous space. We use MICE as a competitor adaptive DoE in Section \ref{sec:experiments}. 

The adaptive sampling algorithm proposed in this paper is based on an improvement metric. At any location, the improvement is defined as the square of the difference between the fitted emulator and the model output at the nearest site. The method, called \emph{variance of improvement for global fit (VIGF)}, uses the variance of improvement as the selection criterion. This criterion can be obtained in closed form and does not require any parameter tuning. Furthermore, its implementation is simpler compared to methods like IMSE and ALC. To encourage broader use, the VIGF implementation is available in the dgpsi R package \cite{dgpsi2024} on CRAN.

At each iteration, a new point is selected where VIGF is the largest, and the emulator is updated with the new data. It is observed that the proposed adaptive approach can efficiently explore the search space with a focus on regions where the model output changes rapidly. The applicability of VIGF is assessed on a couple of benchmark problems and its prediction capability is compared with several sampling strategies. The results suggest that VIGF is the best algorithm in most cases. Section \ref{sec:adaptive_sampling} provides more details on this approach. We first present the basics of GPs in the next section.
\section{Background}
\label{sec:background}
We aim to emulate a deterministic computer code using GPs. The model outputs are produced based on the function $f: \mathcal D \mapsto \mathbb{R}$, where the input space $\mathcal D$ is a compact set in the $d$-dimensional Euclidean space $\mathbb R^d$. We assume that $f$ is computationally expensive and its analytical expression is not known. 
\subsection{GP emulators}
\label{sec:GP}
A GP is a stochastic process that is a collection of random variables, any finite number of which has a multivariate normal distribution \cite{GPML}. GPs are fully specified by their mean and covariance function/kernel that has to be positive semi-definite. Let $\left(Z_0(\bx)\right)_{\bx \in \mathcal{D}}$ represent a GP. In this work, without loss of generality, the GP mean (also known as the trend function) is assumed to be a constant: $\Exp\left[Z_0(\bx) \right] = \mu$. The covariance function $k_0: \mathcal{D} \times \mathcal{D} \mapsto \mathbb R$ is given by
\begin{equation}
	k_0\left(\bx, \bx^\prime\right) = \Cov\left(Z_0(\bx) , Z_0(\bx^\prime) \right) = \sigma^2 \Cor\left(Z_0(\bx) , Z_0(\bx^\prime) \right) , \, \forall \bx, \bx^\prime \in \mathcal{D} , 
\end{equation}
where $\sigma^2$ is the process variance and regulates the scale of the amplitude of $Z_0(\bx)$. Covariance functions play a key role in the GP modelling as most assumptions about the form of $f$ are encoded through them. There is a vast variety of kernels, a list of which can be found in \cite{GPML, sacks1989}. 

Now suppose that the model is evaluated at $n$ design locations $\mathbf{X}_n=\left(\bx_1, \ldots, \bx_n \right)^\top$ with the corresponding outputs $\mathbf{y}_n =\left(f(\bx_1), \dots, f(\bx_n) \right)^\top$. Together, $\mathbf{X}_n$ and $\mathbf{y}_n$ form the training set $\mathcal{A} = \{\mathbf{X}_n,  \mathbf{y}_n\}$. The predictive/posterior distribution of $Z_0(\bx)$, denoted by $Z_n(\bx)$, can be calculated using Bayes' theorem: $Z_n(\bx) = Z_0(\bx) \mid \mathcal{A}$. The posterior process $Z_n(\bx)$ is still a GP with mean and covariance given by \cite{GPML}
\begin{align} 
	\label{E:kriging_mean}
	m_n(\bx) = \Exp\left[Z_n(\bx)\right] &= \hat \mu + \mathbf{k}(\bx)^\top \mathbf{K}^{-1}\left(\mathbf{y}_n -  \hat \mu \mathbf{1} \right) , \\
	\label{E:kriging_cov}
	\nonumber k_n\left(\bx, \bx^\prime\right) = \Cov\left(Z_n(\bx), Z_n(\bx^\prime)\right) &=  k_0(\bx, \bx^\prime)  - \mathbf{k}(\mathbf{x})^\top \mathbf{K}^{-1}\textbf{k}(\mathbf{x^\prime}) \\ 
	& + \frac{\left(1 - \mathbf{k}(\bx)^\top\mathbf{K}^{-1}\mathbf{1} \right)\left(1 - \mathbf{k}(\bx^\prime)^\top\mathbf{K}^{-1}\mathbf{1}\right)}{\mathbf{1}^\top\mathbf{K}^{-1}\mathbf{1}} .
\end{align}
Here, $\textbf{k}(\textbf{x})= \left(k_0(\bx, \bx_1), \dots, k_0(\bx, \bx_n)\right)^\top$, and $\mathbf{1}$ stands for a vector of ones, and $\textbf{K}$ is an $n \times n$ covariance matrix with entries: $\textbf{K}_{ij} = k_0(\bx_i, \bx_j)$, for $1\leq i, j \leq n$. The estimate of $\mu$ is obtained as $\hat \mu = \left(\mathbf{1}^\top\mathbf{K}^{-1}\mathbf{1} \right)^{-1}\mathbf{1}^\top\mathbf{K}^{-1}\mathbf{y}_n$ \cite{GPML}. The predictive variance, i.e., $s^2_n(\bx) = k_n(\bx, \bx)$, is equivalent to the MSE criterion 
\begin{equation}
	MSE(\bx) = \Exp \left[\left( Z_n(\bx) - m_n(\bx) \right)^2 \right]  = s^2_n(\bx) . 
	\label{MSE_criterion}
\end{equation}
It is worth mentioning that the predictive mean interpolates the observations at the experimental designs and the predictive variance is zero there. Also, we have $Z_n(\bx) \sim \mathcal{N} \left(m_n(\bx), s^2_n(\bx)\right), \forall \bx \in \mathcal{D}$.
\subsection{Multi-fidelity emulation}
\label{sec:HierarchicalKriging}
There are many situations across computational science and engineering where multiple computer models are available to estimate the same quantity of interest with different levels of fidelity \cite{peherstorfer2018}. This section deals with the extension of GPs to the multi-fidelity (or multi-level) emulation scheme, where the high-fidelity simulation  ($f$) is predicted with the assistance of lower fidelity model(s). In the multi-fidelity framework, a trade-off between the computational cost and precision is made. On the one hand, the high-fidelity simulator provides reliable results but is computationally demanding. On the other hand, the fast low-fidelity model yields predictions with limited accuracy. In this work, we consider the case of two levels of fidelity, where in addition to the high-fidelity simulator, there is a lower fidelity model ($\tilde f$) with considerably less computational complexity. We tackle this problem via the \emph{hierarchical kriging} (HK) method proposed by Han and G\"{o}rtz \cite{han2012}. With HK, there is no need to model the cross-correlation between the low and high-fidelity simulators, contrary to the conventional cokriging-based approach \cite{kennedy2000}. It is also reported that HK provides a reasonable estimation of the prediction uncertainty. Furthermore, HK can be implemented with some small modifications to an existing (modular) GP code \cite{abdallah2019}. The idea behind the HK approach is that the low-fidelity function is used to capture the general trend of $f$. To do this, the low-fidelity model is emulated and the predictive mean is used as the trend function in the GP for the high-fidelity simulator. The steps are explained in more detail below.

Suppose that the low-fidelity model is evaluated at $N$ points (normally $N > n$) and the training set $\{\mathbf{\tilde X}_{N}, \mathbf{\tilde y}_{N}\}$ is created. Here, the former contains sample locations and the latter consists of the corresponding low-fidelity observations: $\mathbf{\tilde y}_{N} = \tilde f \left(\mathbf{\tilde X}_{N}\right)$. Let $\tilde m_{N}(\bx)$ and $\tilde s^2_{N}(\bx)$ be the predictive mean and variance associated with the emulation of $\tilde f$, respectively. They are computed in a similar fashion as Equations (\ref{E:kriging_mean}) and (\ref{MSE_criterion}). In the HK emulator, denoted by $Z_{0, HK}(\bx)$, the trend function is given by $\beta \tilde m_{N}(\bx)$ where the scaling factor $\beta$ regulates the correlation between the low and high-fidelity emulators \cite{HK_zhang2018}. Using the training set $\mathcal{A}$, the predictive distribution of the HK emulator at $\bx\in\mathcal{D}$ is Gaussian characterised by
\begin{align} 
	\label{HK_kriging_mean}
	m_{n, HK}(\bx) &= \hat\beta \tilde m_{N}(\bx) + \mathbf{k}(\bx)^\top \mathbf{K}^{-1} \left(\mathbf{y}_n -  \hat \beta \mathbf{F} \right) , \\
	s^2_{n, HK}(\bx) &= k_0(\bx, \bx)  - \mathbf{k}(\mathbf{x})^\top \mathbf{K}^{-1}\textbf{k}(\mathbf{x^\prime}) + \frac{\left(\tilde m_{N}(\bx) - \mathbf{k}(\mathbf{x})^\top\mathbf{K}^{-1}\mathbf{F} \right)^2}{\mathbf{F}^\top\mathbf{K}^{-1}\mathbf{F}} ,
	\label{HK_kriging_var}
\end{align}
wherein $\mathbf{F} = \left(\tilde m_{N}(\bx_1), \ldots, \tilde m_{N}(\bx_n)\right)^\top$, and $\hat\beta = \left(\mathbf{F}^\top\mathbf{K}^{-1}\mathbf{F} \right)^{-1}\mathbf{F}^\top\mathbf{K}^{-1}\mathbf{y}_n$ is the estimate of $\beta$ \cite{han2012}. 
\section{Proposed adaptive design criterion}
\label{sec:adaptive_sampling}
Our proposed adaptive sampling algorithm relies on the VIGF criterion by which the initial design is extended sequentially to improve the emulator. At each iteration, a new site is selected where VIGF reaches its maximum and the model is evaluated there to get the corresponding response. This new data is then added to the current training set and the emulator is updated accordingly. Algorithm \ref{proposed_alg} outlines the steps of the proposed method. 
\begin{algorithm}[htpb]
	\caption{Proposed adaptive DoE algorithm}
	\label{proposed_alg}
	Input: training set $\mathcal{A} = \{\mathbf{X}_n,  \mathbf{y}_n\}$, and prior GP $\left(Z_0(\bx)\right)_{\bx \in \mathcal{D}}$
	\begin{algorithmic}[1]
		\STATE Compute the posterior GP $Z_n(\bx) = Z_0(\bx) \mid \mathcal{A}$
		\WHILE{\NOT stop}
		\STATE Compute $\bx_{n+1} = \underset{\bx \in \mathcal{D}}{\arg\!\max}~ VIGF(\bx)$ 
		\STATE Evaluate $f$ at $\bx_{n+1}$ to have $f(\bx_{n+1})$
		\STATE Set $\mathbf{X}_{n+1} = \mathbf{X}_n \cup \left\{\bx_{n+1} \right\}$ and $\mathbf{y}_{n+1} = \mathbf{y}_n \cup \left\{f(\bx_{n+1}) \right\}$
		\STATE Update $Z_n(\bx)$ with $\left( \bx_{n+1}, y_{n+1} \right)$
		\STATE  $n \gets n+1$
		\ENDWHILE
	\end{algorithmic}
\end{algorithm}

The building block of VIGF is the \emph{improvement} function defined as \cite{lam2008}    
\begin{equation}
	\mathcal{I}(\bx) = \left( Z_n(\bx) - f(\bx_i^*) \right)^2 ,
	\label{Improve_fun}
\end{equation} 
with $\bx_i^*$ being the design point closest (in Euclidean distance) to $\bx$. The variance of $\mathcal{I}(\bx)$ can be calculated given that $\mathcal{I}(\bx)/s^2_n(\bx)$ has a noncentral chi-square distribution \cite{mohammadi2022}. It is characterised by 
\begin{equation}
	\mathcal{I}(\bx)/s^2_n(\bx) \sim \chi^{\prime}{^2} \left(\kappa = 1, \lambda = \left( \frac{m_n(\bx)  - f(\bx_i^*)}{s_n(\bx)} \right)^2 \right) ,
	\label{noncentral_chi}
\end{equation}
in which $\kappa$ and $\lambda$ represent the degree of freedom and noncentrality parameter, respectively. We note that the variance of the above distribution is:
\begin{equation}
	\Var \left(\mathcal{I}(\bx)/s^2_n(\bx) \right) = 4\lambda + 2\kappa = 4 \left( \frac{m_n(\bx) - f(\bx_i^*)}{s_n(\bx)} \right)^2  + 2 .
	\label{noncentral_chi_var}
\end{equation}
Now, if both sides of Equation (\ref{noncentral_chi_var}) are multiplied by $s^4_n(\bx)$, we reach the VIGF expression
\begin{equation}
	VIGF(\bx) = \Var \left(\mathcal{I}(\bx)\right) = 4 s_n^2(\bx) \left( m_n(\bx) - f(\bx_i^*) \right)^2 + 2s_n^4(\bx) .
	\label{VIGF_criterion}
\end{equation}
 The terms $s_n^2(\bx)$ and $\left( m_n(\bx) - f(\bx_i^*) \right)^2$, which are called the variance and squared \emph{bias} (with respect to $f(\bx_i^*)$) in conventional statistics, account for exploration and exploitation, respectively. The former quantifies uncertainty in the GP prediction at $\bx$, and grows away from the training data. The latter captures the local behaviour of $f$, and gets large if the difference between $m_n(\bx)$ and $f(\bx_i^*)$ increases. Due to the (multiplicative) interaction between the exploration and exploitation terms in Equation (\ref{VIGF_criterion}), VIGF augments in regions where both components are critical.

\begin{figure}[htpb] 
	\centering
	\includegraphics[width=0.49\textwidth]{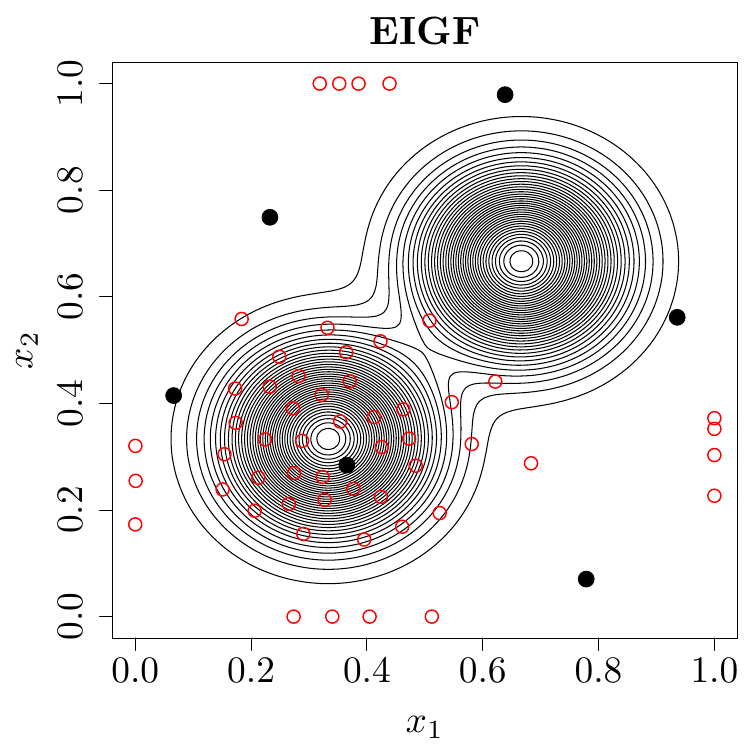}
	\includegraphics[width=0.49\textwidth]{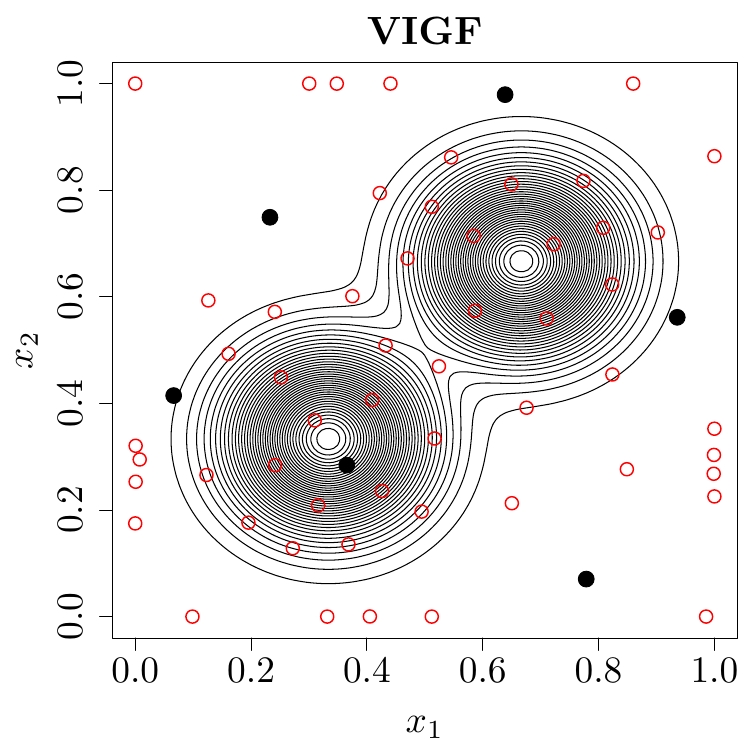}  
	\caption{The sampling behaviour of EIGF (left) and VIGF (right) on a 2-dimensional problem which is a sum of two Gaussian functions. The black dots represent the initial design locations and red circles are the samples selected adaptively. EIGF takes many samples around the optimum $(1/3, 1/3)^\top$. VIGF has a more global search due to the interaction between the exploration and exploitation components.}
	\label{EIGF_vs_VIGF}
\end{figure}

It is notable that Lam \cite{lam2008} proposed \emph{expected improvement for global fit (EIGF)} as an adaptive sampling criterion. It is the expected value of $\mathcal{I}(\bx)$ given by
\begin{equation}
	EIGF(\bx) = \Exp \left[\mathcal{I}(\bx)\right] = \left( m_n(\bx) - f(\bx_i^*) \right)^2 + s_n^2(\bx) .
	\label{EIGF_criterion}
\end{equation}
The main disadvantage of EIGF is that it tends towards local exploitation, and is susceptible to get stuck in an optimum \cite{beck2016, mohammadi2022}. The reason is that  when $\bx_i^*$ is near an optimum, the exploitation term $\left( m_n(\bx) - f(\bx_i^*) \right)^2$ grows while the predictive variance $s_n^2(\bx)$ is not significant. As a consequence, EIGF tends to add new samples near the optimum. More rigorously, since the partial derivatives of EIGF with respect to $\lvert m_n(\bx) - f(\bx_i^*)\rvert$ is $2\big \lvert m_n(\bx) - f(\bx_i^*) \big \rvert$, it is not easy for EIGF to leave the basin of attraction when the exploitation term is massive. VIGF does not suffer from this problem though due to the multiplicative relation between the exploration exploitation components, see Equation (\ref{VIGF_criterion}). The partial derivatives of VIGF with respect to $\lvert m_n(\bx) - f(\bx_i^*)\rvert$ obeys $8 s_n^2(\bx) \big \lvert m_n(\bx) - f(\bx_i^*) \big \rvert$ in which the exploitation term is multiplied by $8s_n^2(\bx)$. As a result, VIGF can escape the basin of attraction and explore other areas within the input space. To shed more light on this, Figure \ref{EIGF_vs_VIGF} is used to visualise the sampling behaviour of EIGF (left) and VIGF (right) when exploitation plays a critical role. The underlying function $f$, is a sum of two Gaussian functions. The black dots are the initial design, and the red circles represent the adaptive samples generated by the two approaches. As can be seen, EIGF overexploits the area around the centre of spike at $(1/3, 1/3)^\top$ and puts a lot of samples there. In VIGF, the balance between exploration and exploitation is more efficient and the search is more global than EIGF with.
\paragraph{Extension to batch mode} The VIGF criterion is limited to sampling only one point per iteration. However, batch-mode sampling allows for the selection of multiple inputs for evaluation in each iteration. This method is particularly useful when parallel computing resources are available, as the evaluation of the points in the bath can be distributed across multiple cores or processors. This parallel processing significantly reduces the wall-clock time, i.e., the time the user waits for the computations to complete. Here, we propose the batch version of the VIGF criterion, referred to as \emph{pseudo} VIGF (PVIGF). It is obtained by multiplying VIGF by a repulsion (or influence) function (RF) defined as \cite{zhan2017} 
\begin{equation}
	RF(\bx; \bx_u) = 1 - \Cor\left(Z_n(\bx), Z_n(\bx_u) \right) ,
\end{equation} 
with $\bx_u$ being an updating point. The RF function is zero at $\bx_u$ and tends to one as $\lVert \bx - \bx_u \rVert \to\infty$.

\begin{figure}[htpb] 
	\centering
	\includegraphics[width=0.49\textwidth]{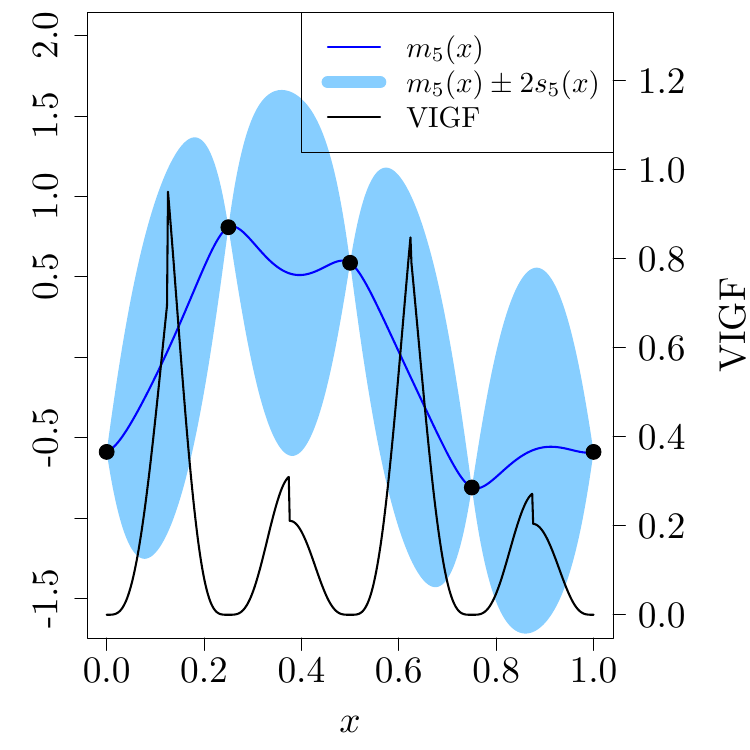}
	\includegraphics[width=0.49\textwidth]{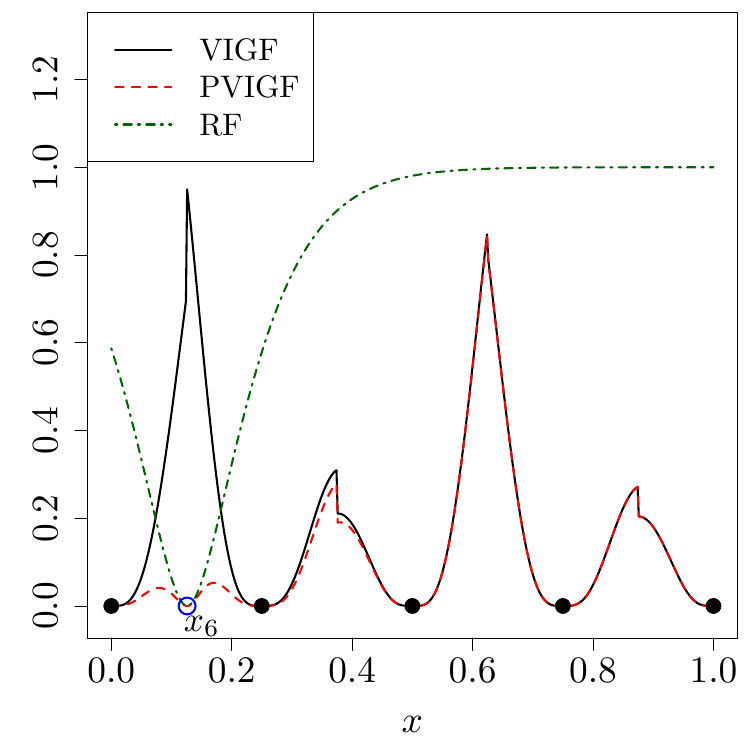}  
	\caption{Left: A GP fitted to the five training data (black dots), and the corresponding VIGF function (right scale). Right: PVIGF (dashed red) is computed by multiplying VIGF by the RF function (dash-dotted green) with the updating point $x_6$ (blue circle). This process illustrates how PVIGF is updated by adding $x_6$ to the existing design without requiring the evaluation of $f(x_6)$, which would be costly to compute.}
	\label{PVIGF}
\end{figure}

 With the RF function one can add new design locations (i.e., updating points) to the current DoE without the need to evaluate the expensive function $f$. In essence, PVIGF approximates VIGF that would be updated by incorporating the new candidate point (i.e., $(\bx_u, f(\bx_u))$) into the current DoE. This is illustrated in Figure \ref{PVIGF} where the left panel shows a GP fitted to five design points (black dots), and the corresponding VIGF criterion (right scale). In the right panel, the new location $x_6$ (blue circle) at which VIGF is maximum is added to the existing design. PVIGF (dashed red) is obtained by multiplying $VIGF(x)$ by $RF(x; x_6)$ (dash-dotted green). Note that, this process does not require knowledge of the actual value of $f(x_6)$, which means that new points can be incorporated without evaluating the expensive function.

More rigorously, let $q > 1$ represent the number of sites to be added to $\mathbf{X}_n$ at each iteration using batch sampling. In this approach, the first new input $\bx_{n+1}$ is selected by maximising VIGF as explained earlier. The subsequent sampling locations are determined by maximising PVIGF, which is calculated by multiplying the initial VIGF criterion by the RF functions associated with the batch members selected so far. This relationship can be mathematically expressed as follows:
\begin{align*}
	\bx_{n+2} &= \underset{\bx \in \mathcal{D}}{\arg\!\max}~ PVIGF(\bx) = RF(\bx; \bx_{n+1}) VIGF(\bx) , \\
	\bx_{n+3} &= \underset{\bx \in \mathcal{D}}{\arg\!\max}~ PVIGF(\bx) =  RF(\bx; \bx_{n+1}) RF(\bx; \bx_{n+2}) VIGF(\bx), \\
	\vdots&   \hspace{2cm}  \vdots \hspace{4cm}  \vdots \\
	\bx_{n+q} &= \underset{\bx \in \mathcal{D}}{\arg\!\max}~ PVIGF(\bx) =  RF(\bx; \bx_{n+1}) \ldots RF(\bx; \bx_{n+q-1}) VIGF(\bx) .
\end{align*} 
The pseudocode of the batch mode sampling is presented in Algorithm \ref{batch_alg}.
\begin{algorithm}[htpb]
	\caption{Choosing a batch of $q > 1$ locations at each iteration} 
	\label{batch_alg}
	Input: Posterior GP $Z_n(\bx)$ constructed based on $\mathcal{A} = \{\mathbf{X}_n,  \mathbf{y}_n\}$
	\begin{algorithmic}[1]
		\STATE Create $\text{VIGF}(\bx)$ using Equation (\ref{VIGF_criterion})
		\STATE $\bx_{n+1} = \underset{\bx \in \mathcal{D}}{\arg\!\max}~ VIGF(\bx)$ 
		\STATE $PVIGF(\bx) \gets RF(\bx; \bx_{n+1}) VIGF(\bx)$
		\FOR {$j=2$ to $q$}
		\STATE $\bx_{n+j} = \underset{\bx \in \mathcal{D}}{\arg\!\max}~ PVIGF(\bx)$ 
		\STATE $PVIGF(\bx) \gets RF(\bx; \bx_{n+j})PVIGF(\bx)$
		\ENDFOR
	\end{algorithmic}
	Output: $\bx_{n+1}, \bx_{n+2}, \ldots, \bx_{n+q}$
\end{algorithm}
\paragraph{Extension to multi-fidelity modelling} Now, we investigate the potential of VIGF for multi-fidelity modelling applications although this is not the main focus of the current paper. To this end, we extend the proposed adaptive DoE to a two-level fidelity case in which a lower fidelity model contributes to the prediction of the high-fidelity simulator $f$. We tackle this problem by adapting the \emph{variable-fidelity} method introduced in \cite{HK_zhang2018} relying on the HK model. The advantage of this approach is that the uncertainty due to the lack of a low-fidelity observation is analytically available, which may not be easily attainable in other approaches. As a result, we can compute the mean and variance of the improvement function in closed form. This is further explained below.

It is shown that the predictive variance of the HK emulator associated with the lack of a low-fidelity sample at $\bx\in\mathcal{D}$ is equal to $\beta \tilde s^2_{N}(\bx)$ \cite{HK_zhang2018}. Hence, the prediction uncertainty of the HK model can be expressed as a function of both the input values and the fidelity level as below
\begin{equation}
	s^2_{n, HK}(\bx, L) = 
	\begin{cases} 
		\beta^2\tilde s^2_{N}(\bx) & \text{if } L = 1\\ 
		s^2_{n, HK}(\bx) &\text{if } L = 2  .
	\end{cases}  
\label{MSE_HK_criterion}
\end{equation}
Here, $L$ indicates the fidelity level with $L = 1$ for the low-fidelity model and $L = 2$ for the high-fidelity one. In this setting, the prediction of the HK model and the improvement function at any point $\bx$ are
\begin{align}
	&Z_{n, HK}(\bx, L) \sim \mathcal{N} \left(m_{n, HK}(\bx), s^2_{n, HK}(\bx, L)\right) , \\
	&\mathcal{I}_{HK}(\bx, L) = \left(Z_{n, HK}(\bx, L) - f(\bx_i^*) \right)^2 ,\quad L = 1, 2 . 
	\label{Improve_fun_HK}
\end{align}
Note that $\mathcal{I}_{HK}(\bx, L)$ is a measure of the distance between $Z_{n, HK}(\bx, L)$ and the high-fidelity model output at $\bx_i^*$, i.e., the closest design location to $\bx$. 

Now, the adaptive sampling criteria EIGF and VIGF can be easily extended to the two-level fidelity case. The new criteria, denoted by \text{$EIGF_{HK}$} and \text{$VIGF_{HK}$}, are obtained if $m_n(\bx)$ and $s^2_n(\bx)$ in Equations (\ref{VIGF_criterion}) and (\ref{EIGF_criterion}) are replaced with $m_{n, HK}(\bx)$ (Equation (\ref{HK_kriging_mean})) and $s^2_{n, HK}(\bx, L)$ (Equation (\ref{MSE_HK_criterion})), respectively. For example, \text{$VIGF_{HK}$} takes the following form
\begin{equation}
	VIGF_{HK}(\bx, L) =  4 s^2_{n, HK}\left(\bx, L\right) \left( m_{n, HK}(\bx) - f(\bx_i^*) \right)^2 + 2s^4_{n, HK}\left(\bx, L\right) .
	\label{VIGF_HK_criterion}
\end{equation}
Note that the MSE criterion in a two-level fidelity model (denoted by \text{$MSE_{HK}$}) is expressed by Equation (\ref{MSE_HK_criterion}). 

Sequential design for multi-fidelity models involves identifying both the location and the fidelity level of the next sample. This problem is often tackled in two separate steps: first selecting the new point, then determining the appropriate fidelity level. However, this approach my lead to a sub-optimal solution. Using the \text{$VIGF_{HK}$} criterion, the next sampling location ($\bx_{n+1}$) and its corresponding fidelity level ($L^\ast$) are determined simultaneously by solving the following optimisation problem:
\begin{equation}
	\bx_{n+1}, L^\ast = \underset{\bx \in \mathcal{D}, \, L = 1, 2}{\arg\!\max}~ VIGF_{HK}(\bx, L) . 
\end{equation}
The same rule applies to \text{$EIGF_{HK}$} and \text{$MSE_{HK}$} when they are employed in a two-level adaptive sampling scheme. It is important to note that multi-fidelity modelling can be performed using methods such as those proposed by Le Gratiet and Garnier \cite{legratiet2014} or Konomi and Karagiannis \cite{konomi2021}. However, these methods often involve more complex recursive co-kriging models or Bayesian frameworks, which can be challenging to implement.
\section{Numerical experiments} 
\label{sec:experiments}
The results of numerical experiments are presented in this section. The applicability of the proposed method is compared with several sampling schemes in both single and two-level fidelity cases on a suite of frequently used analytical test functions. The details are provided below.
\subsection{Test functions and experimental setup}
 Our proposed adaptive DoE is applied to eight well-established benchmark problems \cite{bingham2024} which serve as the ground truth models. They are summarised in Table \ref{tab:test_fun} and their analytical expression is given in Appendix \ref{append:test_fun}. The functions $f_1 - f_6$ are defined on the hypercube $[0, 1]^d$. The problems $f_7$ and $f_8$ are physical models; the former simulates the midpoint voltage of an OTL push-pull circuit and the latter computes the cycle time of a piston within a cylinder. The functions are classified into two groups according to their dimensionality: small-scale ($1 \leq d \leq 4$) and medium-scale ($5 \leq d \leq 8$) \cite{kianifar2020}. In order to assess the accuracy of the emulator, we use the \emph{normalised root mean squared error} (NRMSE) criterion 
\begin{equation}
NRMSE = \frac{\sqrt{\frac{1}{n_t} \sum_{t = 1}^{n_t} \left( m_n(\bx_t) - f(\bx_t) \right)^2 } }{\underset{t = 1:n_t}{\max}\, f(\bx_t) - \underset{t = 1:n_t}{\min}\, f(\bx_t) } ,
\end{equation}
 wherein $\lbrace \left(\bx_t, f(\bx_t) \right) \rbrace_{t = 1}^{t = n_t}$ is the test data set. In our experiments, $n_t = 3000$ and the test points are spread across the domain uniformly. The total number of function evaluations is set to $30 d$ and the size of initial design is $3 d$. The initial training samples are obtained based on the LHS strategy.
\begin{table}[htpb] 
	\caption{Summary of benchmark functions \cite{bingham2024}}
	\centering
	\begin{tabular}{l l c c} 
		 \hline
			& Function name & Dimension ($d$) & Scale \\ [0.5ex] 
		\hline\hline
		$f_1$ & Franke & 2 & Small \\ 
		$f_2$ & Dette \& Pepelyshev (curved) & 3 & Small \\
		$f_3$ & Hartmann & 3 & Small \\
		$f_4$ & Park & 4 & Small \\
		$f_5$ & Friedman & 5 & Medium \\ 
		$f_6$ & Gramacy \& Lee & 6 & Medium \\ 
		$f_7$ & Output transformerless (OTL) circuit & 6 & Medium \\
		$f_8$ & Piston simulation & 7 & Medium \\[1ex] 
		\hline
	\end{tabular}
\label{tab:test_fun}
\end{table}

 To construct GPs, the R package \emph{DiceKriging} \cite{roustant2012} is employed. The covariance kernel is Mat\'ern $3/2$ and the unknown parameters are estimated by maximum likelihood. The Mat\'ern covariance function is a popular choice in the computer experiments literature \cite{gramacy2020}. The optimisation of the adaptive design criteria is performed via the R package \emph{DEoptim} \cite{mullen2011} which implements the differential evolution algorithm \cite{storn1997}. For each function, the sequential sampling methods are run with ten different initial space-filling designs. Thus, we have ten sets of NRMSE for each sampling strategy. The LHS designs are generated using the \texttt{maximinESE\_LHS} function implemented in the R package \emph{DiceDesign} \cite{DiceDesign}. This function employs a stochastic optimisation technique to produce an LHS design based on the maximin criterion.
\subsection{Results and discussion}
\label{sec:results}
\begin{figure}[htpb] 
	\includegraphics[width=0.49\textwidth]{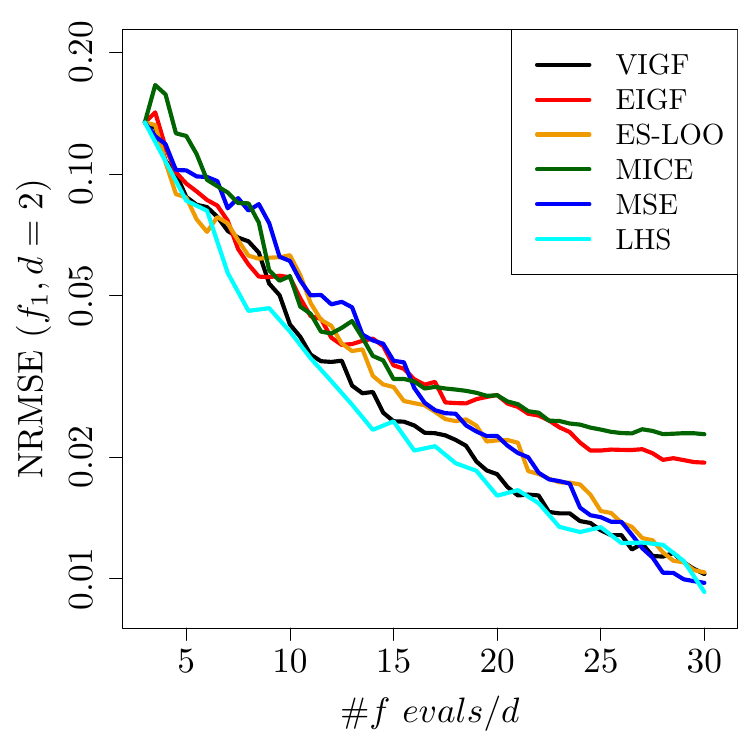}
	\hspace{0.1cm}
	\includegraphics[width=0.49\textwidth]{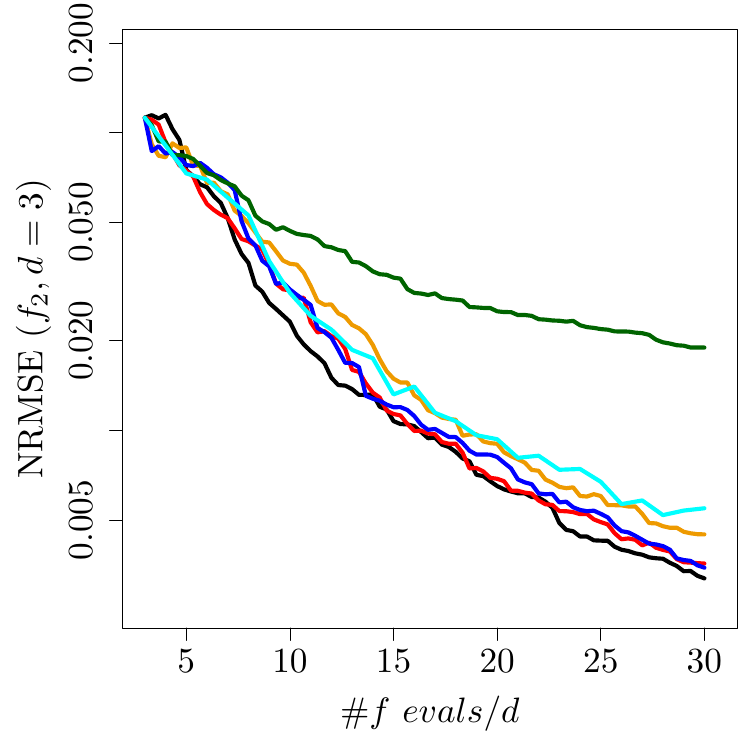} \\
	\includegraphics[width=0.49\textwidth]{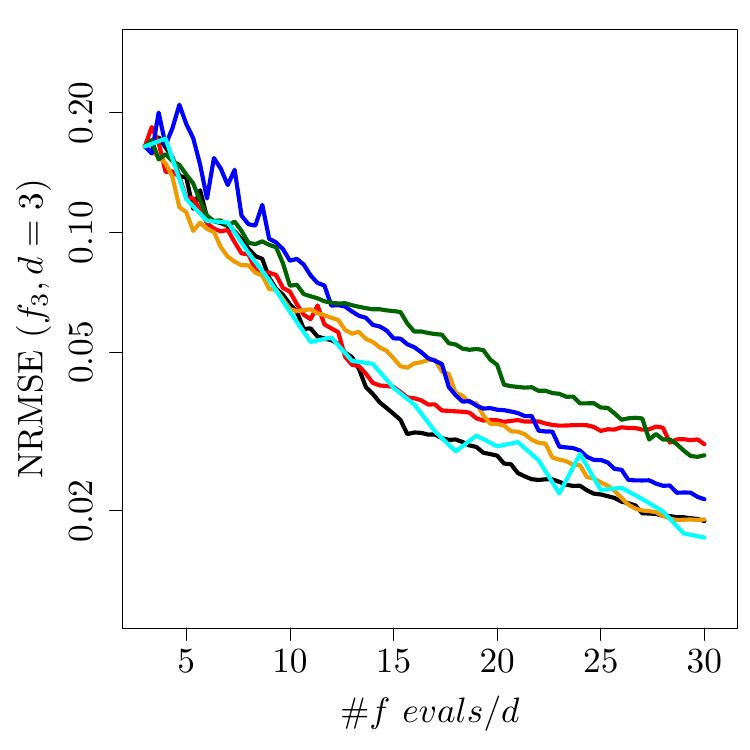}
	\hspace{0.1cm}
	\includegraphics[width=0.49\textwidth]{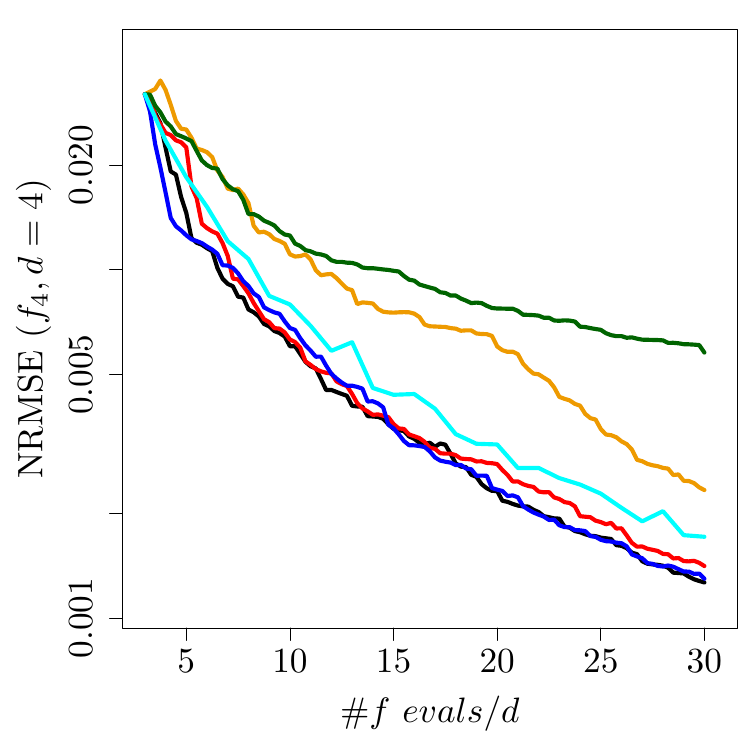}
	\caption{Accuracy results for the small-scale problems, i.e., $1 \leq d \leq 4$. Each method generates ten prediction using ten different initial DoE. Every curve represents the median of ten NRMSEs obtained by VIGF (black), EIGF (red), ES-LOO (orange), MSE (blue), MICE (green), and the one-shot LHS design (cyan). For the latter, the NRMSE is computed for all sample sizes $3d$, $4d, \ldots, 30d$. The $x$-axis shows the number of function calls divided by the problem dimensionality. The $y$-axis is on a logarithmic scale.}
	\label{res_small_scale}
\end{figure}

\begin{figure}[htpb] 
	\includegraphics[width=0.49\textwidth]{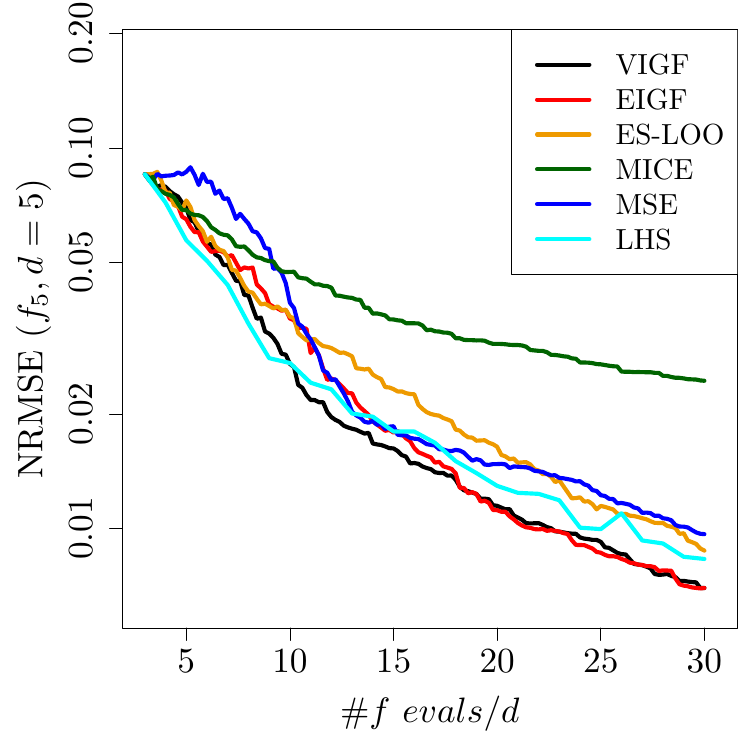} 
	\hspace{0.1cm}
	\includegraphics[width=0.49\textwidth]{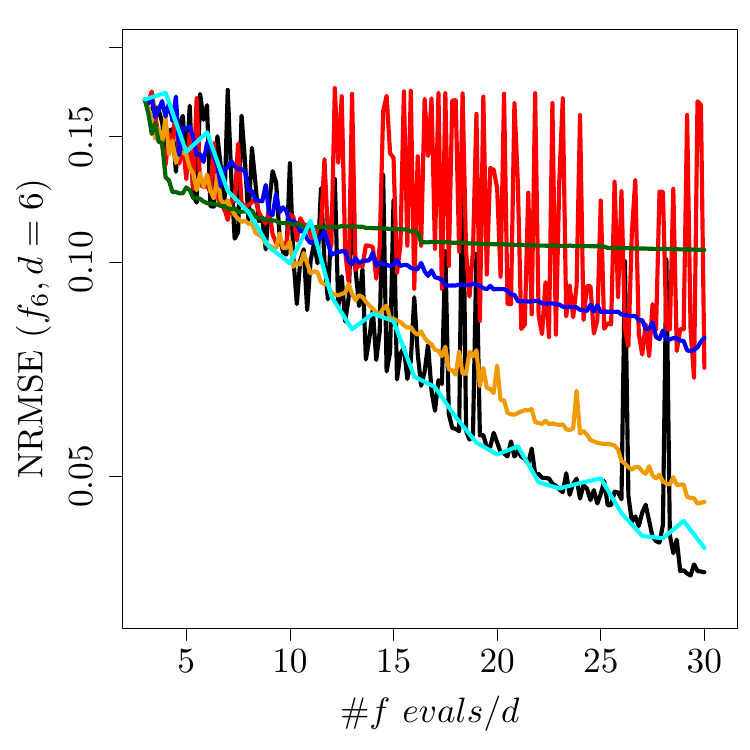} \\
	\includegraphics[width=0.49\textwidth]{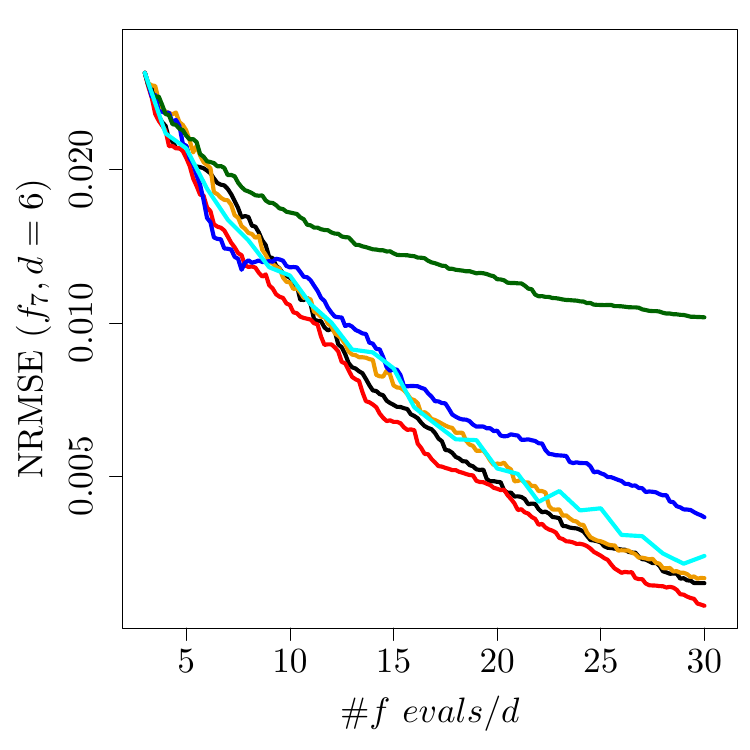}
	\hspace{0.1cm}
	\includegraphics[width=0.49\textwidth]{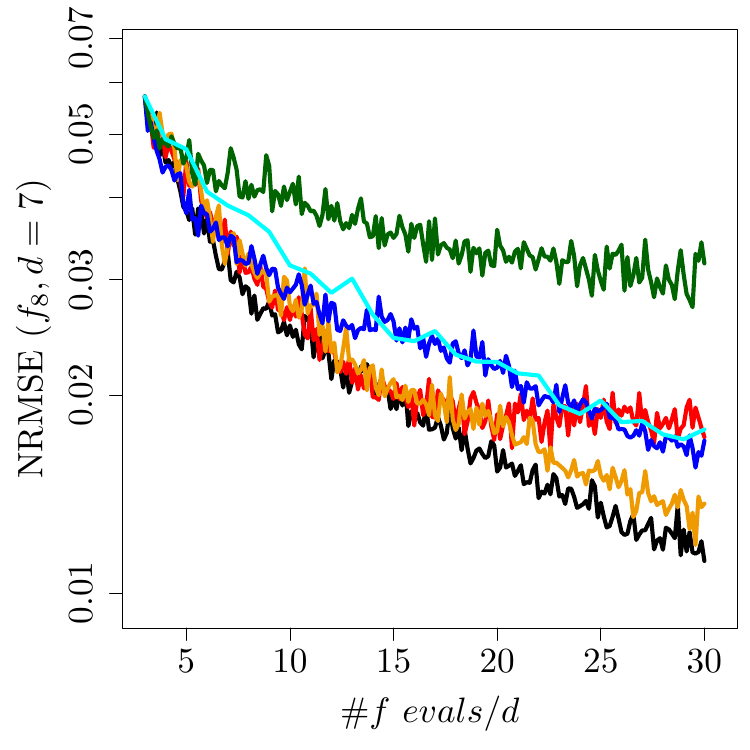} 
	\caption{Accuracy results for the medium-scale problems, i.e., $5 \leq d \leq 8$. Each curve is the median of ten RMSEs associated with the design based on VIGF (black), EIGF (red), ES-LOO (orange), MSE (blue), MICE (green), and the one-shot LHS design (cyan).}
	\label{res_medium_scale}
\end{figure}

The results of our numerical experiments are illustrated in Figures \ref{res_small_scale} (small-scale) and \ref{res_medium_scale} (medium-scale).  Each curve represents the median of ten NRMSEs. The performance of VIGF (black) is compared with a number of adaptive sampling methods, i.e., EIGF (red), ES-LOO (orange), MSE (blue), MICE (green). In addition, the NRMSE based on the one-shot LHS design (cyan) is computed for all sample sizes $3d$, $4d, \ldots, 30d$. The LHS design is also repeated ten times.

\begin{figure}[htpb] 
	\includegraphics[width=0.32\textwidth]{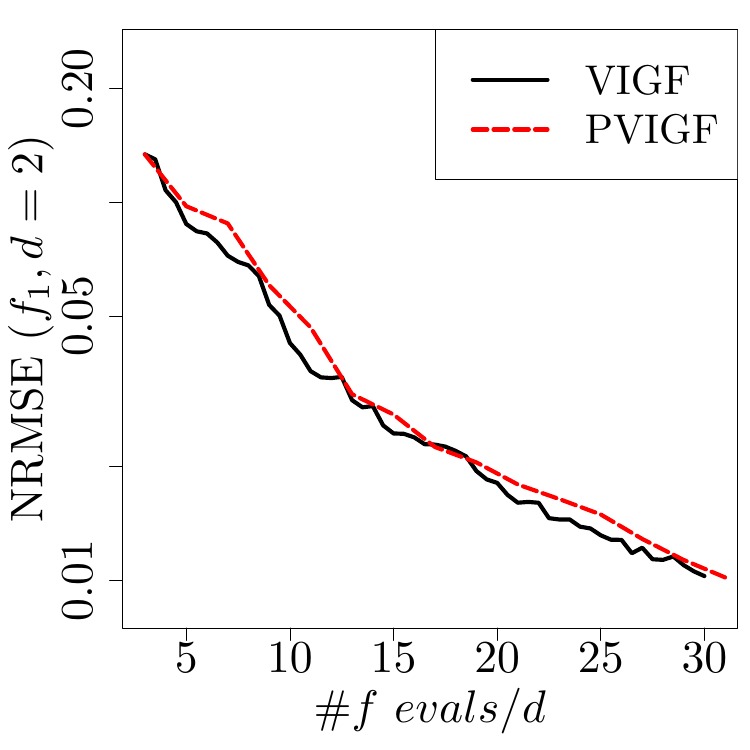}
	\includegraphics[width=0.32\textwidth]{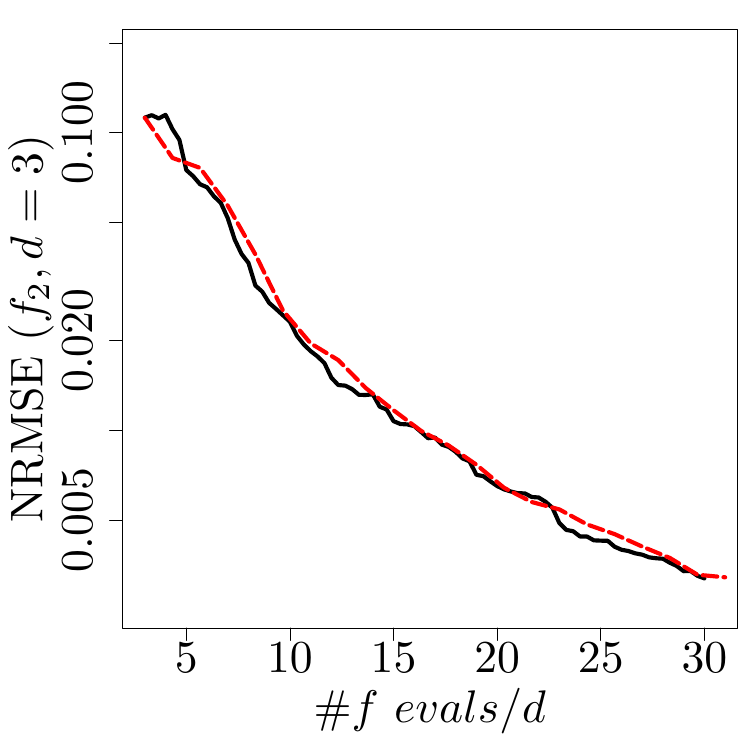}
	\includegraphics[width=0.32\textwidth]{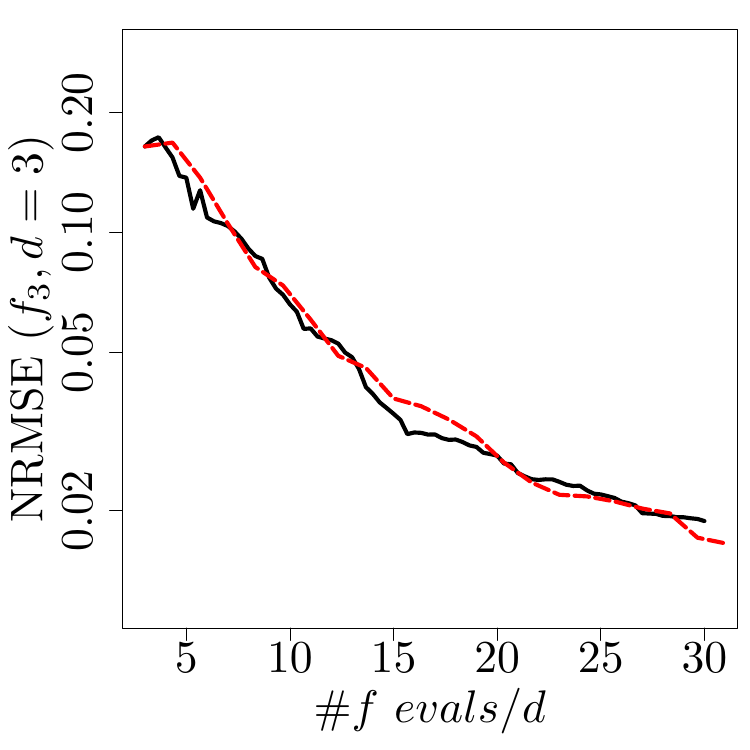} \\
	\includegraphics[width=0.32\textwidth]{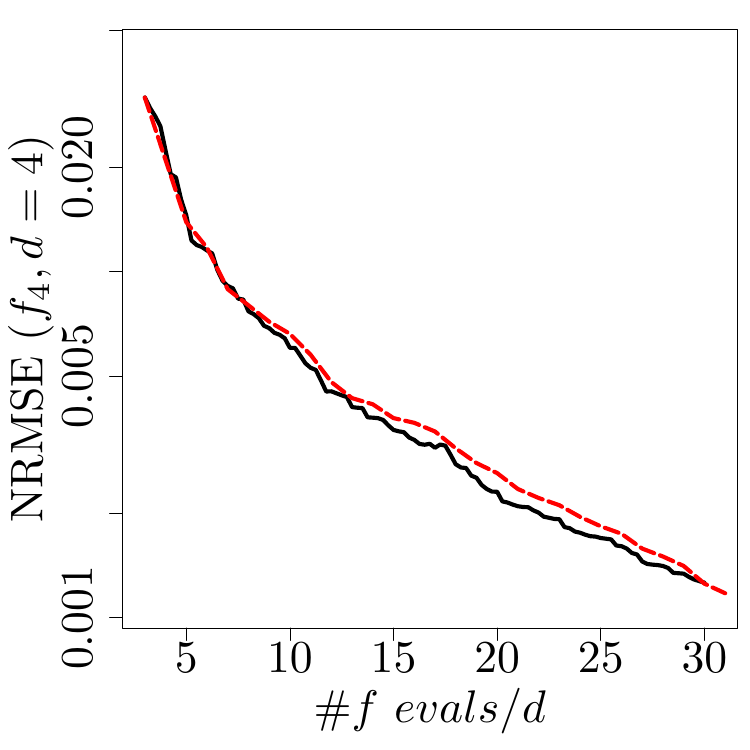} 
	\includegraphics[width=0.32\textwidth]{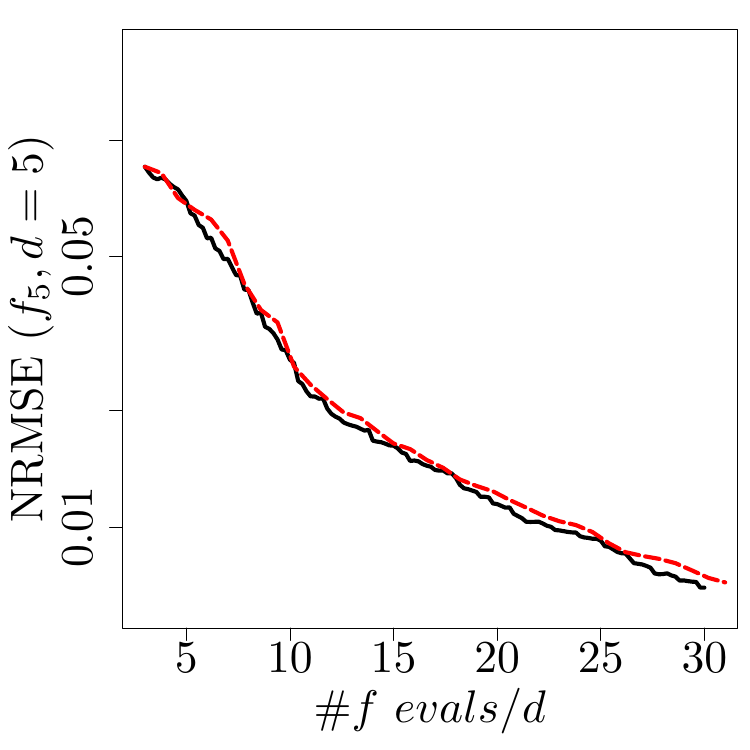} 
	\includegraphics[width=0.32\textwidth]{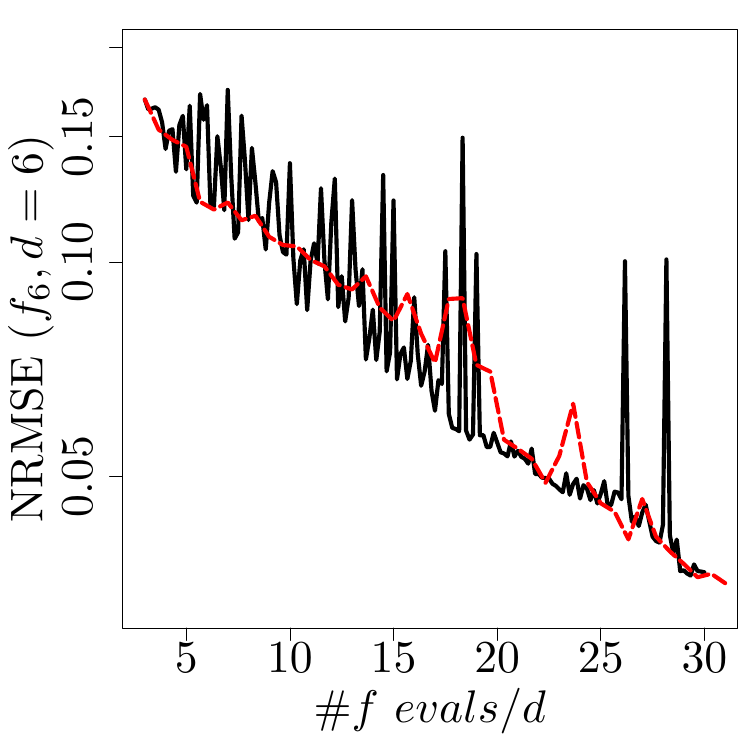} \\
	\includegraphics[width=0.32\textwidth]{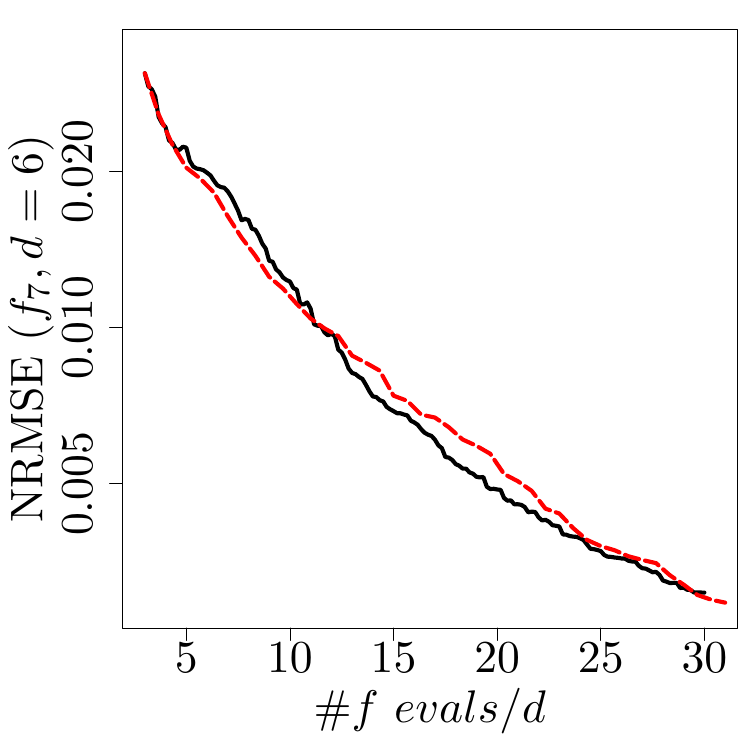}
	\includegraphics[width=0.32\textwidth]{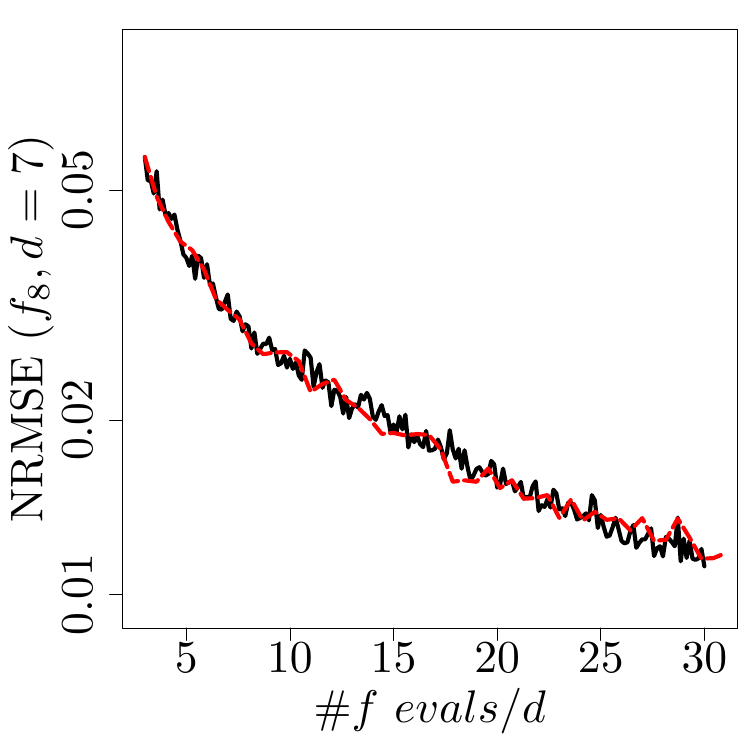} 
	\caption{Accuracy results of VIGF (black) and PVIGF (red dashed) with $q = 4$. As can be seen, VIGF and its batch version have similar performances. The advantage of the batch mode adaptive sampling is that it can save the user time when parallel computing is available.}
	\label{res_PVIGF}
\end{figure}

We observe that the design strategy proposed in this paper outperforms the other approaches in emulating the functions $f_2, f_4, f_5, f_6$, and $f_8$, and its performance is comparable in the remaining cases ($f_1$, $f_3$, and $f_7$). This is due to the ability of VIGF to explore the input space while focusing on interesting regions. The EIGF method is the best algorithm on $f_5$ and $f_7$, however, its capability in emulating the multimodal functions $f_1, f_3$, and $f_6$ is not favourable. It is highly likely that the EIGF criterion gets stuck in an optimum as discussed in Section \ref{sec:adaptive_sampling}. The method based on ES-LOO is never the winning algorithm, its accuracy results are promising yet except for $f_4$. The MSE criterion shows a good capability to predict small-scale problems as it tends to fill the space in a uniform manner, see below for further discussion. However, this is not the case for the medium (and possibly high) dimensional problems where the MSE measure tends to take more samples on the boundaries. The MICE algorithm performs poorly in all the cases due to its implementation on a discrete representation. The performance of the LHS design is comparable to the adaptive sampling approaches in most cases except the physical models, i.e., $f_7$ and $f_8$. The main disadvantage of the one-shot LHS design is that it can waste computational resources due to under/oversampling. We have also applied PVIGF with $q = 4$ to the eight benchmark functions. It is found that PVIGF has a similar performance to VIGF. The results are presented in Figure \ref{res_PVIGF}. 

 The sampling behaviour of the above methods is further investigated using the \emph{discrepancy} criterion \cite{fang2000}. This function assesses the uniformity of a design by measuring how the empirical distribution of the points in the design deviates from the uniform distribution. The $\mathbb{L}^2$-discrepancy of a set $\mathbf{X}_{N^\ast} \subset [0, 1]^d$ consisting of $N^\ast$ points is given by
\begin{equation}
	D_2\left(\mathbf{X}_{N^\ast} \right) =  \left\lbrack \int_{[0, 1]^d} \left\lvert \frac{A \left(\mathbf{X}_{N^\ast}, J_{\bx} \right)}{N^\ast} - \text{Volume} \left(J_{\bx}\right) \right\rvert^2 \,d\bx \right\rbrack^{1/2} ,
\end{equation}
in which $J_{\bx}$ denotes the interval $[0, \bx) = [0, x_1) \times [0, x_2)\times\ldots \times [0, x_d)$, and $A \left(\mathbf{X}_{N^\ast}, J_{\bx} \right)$ returns the number of points of $\mathbf{X}_{N^\ast}$ falling in $J_{\bx}$ \cite{lin2015}.  

\begin{figure}[htpb] 
	\includegraphics[width=0.49\textwidth]{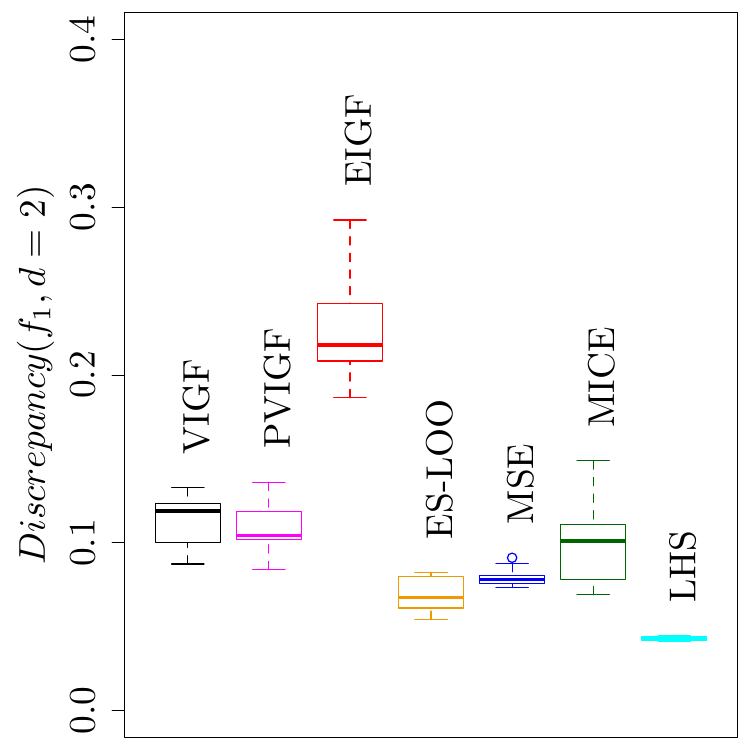}
	\hspace{0.1cm}
	\includegraphics[width=0.49\textwidth]{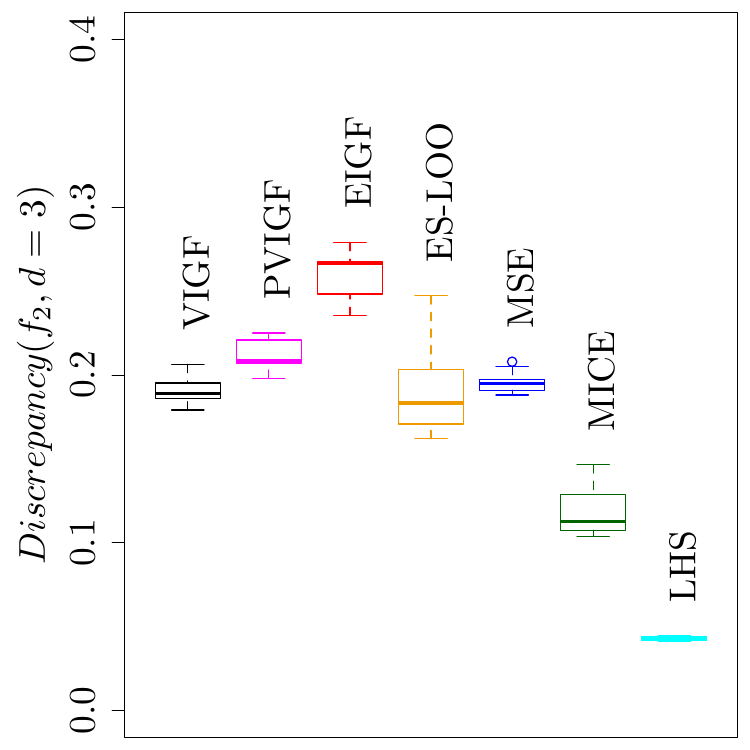} \\
	\includegraphics[width=0.49\textwidth]{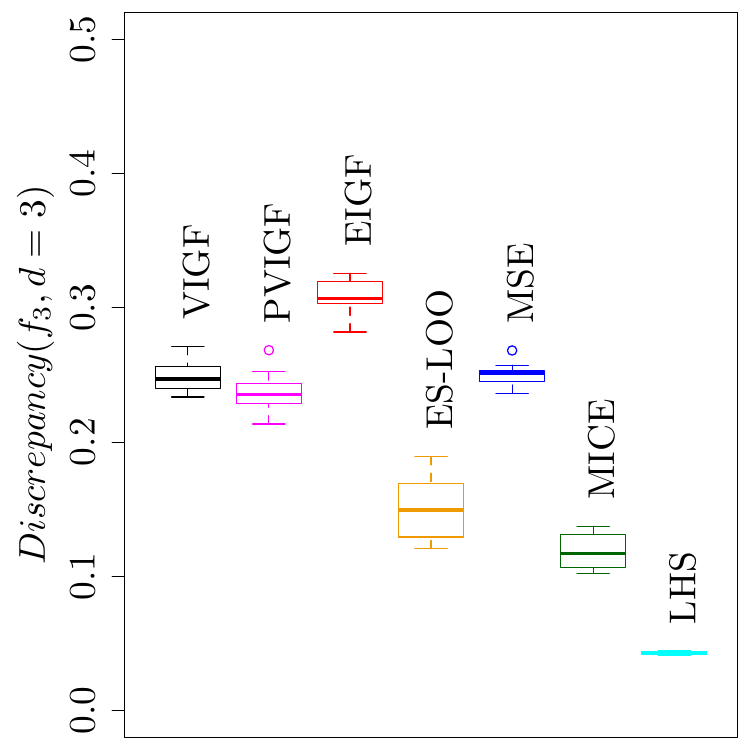}
	\hspace{0.1cm}
	\includegraphics[width=0.49\textwidth]{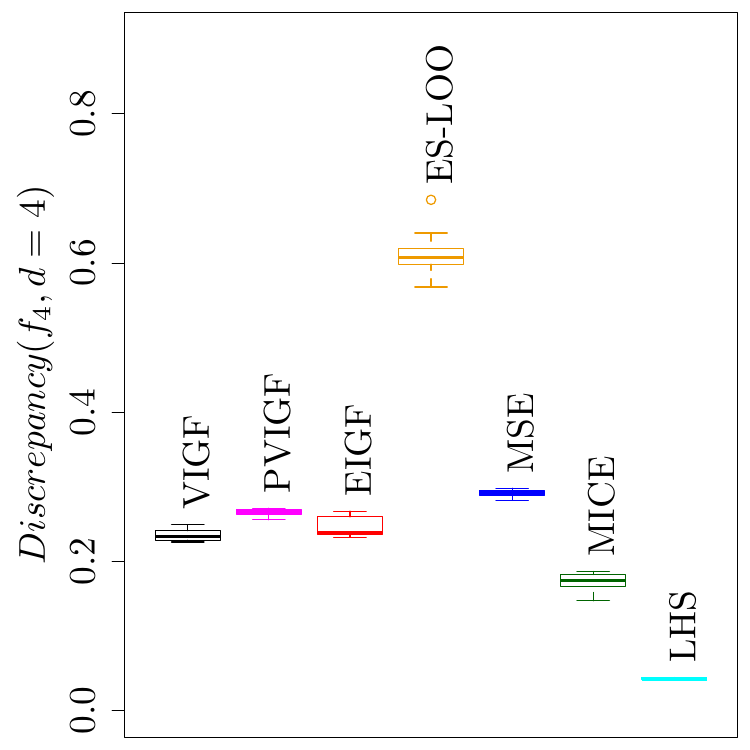}
	\caption{The box plot of the $\mathbb{L}^2$-discrepancy criterion for the designs obtained by VIGF (black), PVIGF (magenta), EIGF (red), ES-LOO (orange), MSE (blue), MICE (green), and LHS (cyan). The discrepancy measures how a given distribution of points differs from the uniform distribution. Accordingly, the LHS design has a very low discrepancy.}
	\label{res1_disc_criter}
\end{figure}

\begin{figure}[htpb]
	\includegraphics[width=0.49\textwidth]{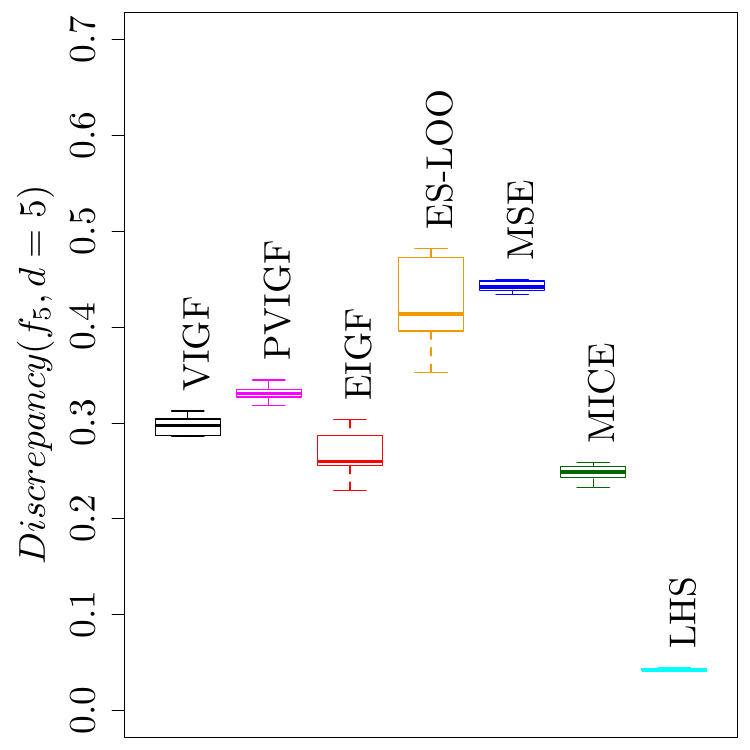} 
	\hspace{0.1cm}
	\includegraphics[width=0.49\textwidth]{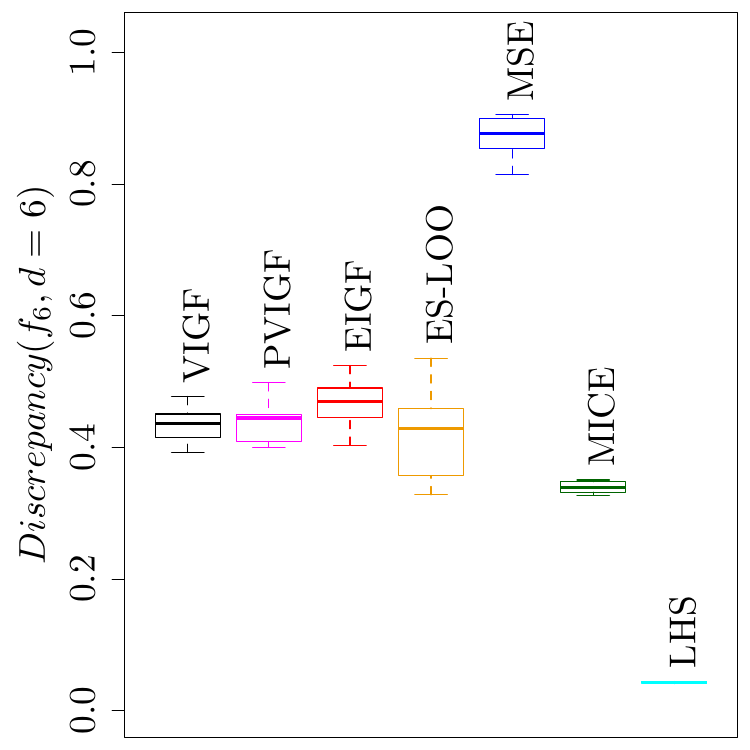} \\
	\includegraphics[width=0.49\textwidth]{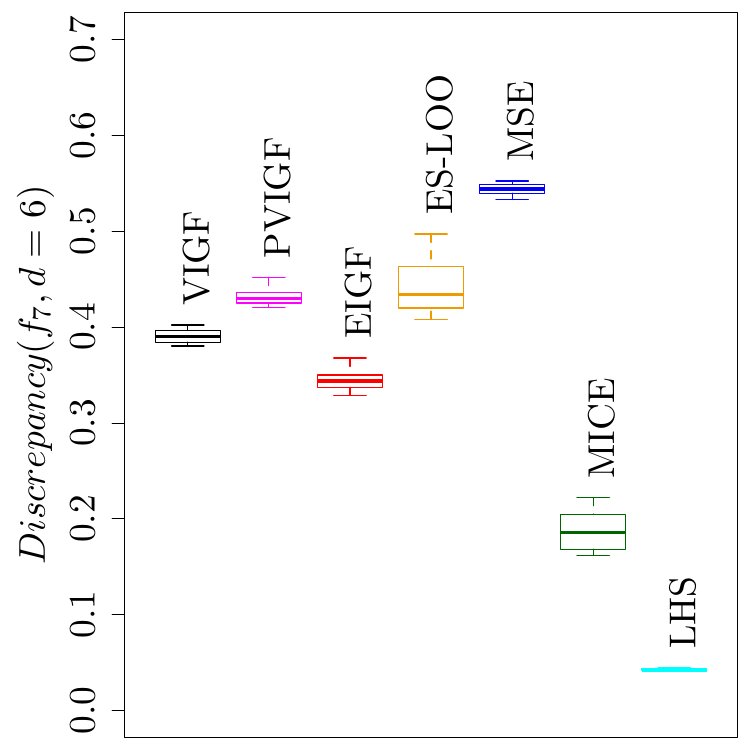}
	\hspace{0.1cm}
	\includegraphics[width=0.49\textwidth]{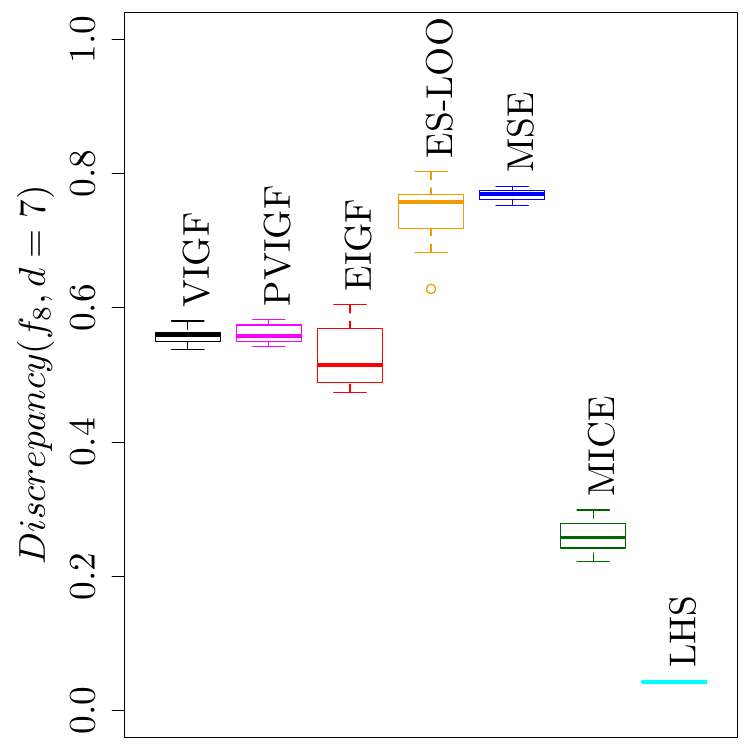} 
	\caption{The box plot of the $\mathbb{L}^2$-discrepancy criterion for the methods based on VIGF (black), PVIGF (magenta), EIGF (red), ES-LOO (orange), MSE (blue), MICE (green), and LHS (cyan). The sampling scheme based on MSE leads to a large discrepancy in the medium-scale problems as many points are samples on their boundaries.}
\label{res2_disc_criter}
\end{figure}

We have computed the $\mathbb{L}^2$-discrepancy for all the sampling strategies used in this study. Figures \ref{res1_disc_criter} (small-scale) and \ref{res2_disc_criter} (medium-scale) display the box plot of the discrepancy results. As expected, the discrepancy of the LHS design is close to zero with a small variance. The method based on VIGF has neither the lowest nor highest discrepancy among the adaptive sampling approaches. This is due to the characteristic of VIGF that tends to fill the space with a focus on interesting regions. The discrepancy of PVIGF is close to that of VIGF. The discrepancy associated with the EIGF criterion is large in the problems $f_1, f_3$, and $f_6$ where this criterion performs poorly. This can be explained by the tendency of EIGF towards local exploitation in those functions. It is observed that the design produced by the MICE algorithm has the smallest discrepancy after the LHS method. The discrepancy of MSE is significantly larger in the medium-scale functions than small-scale ones. The reason is that in the medium-scale functions many points are selected on their boundaries where the GP predictive variance is enormous.

\subsection{Two-level fidelity results}
\label{sec:multi-fidelity}
We assess the prediction capability of the proposed \text{$VIGF_{HK}$} algorithm in the two-level fidelity case and compare the results with \text{$EIGF_{HK}$} and \text{$MSE_{HK}$}. To do this, we consider four test examples that are commonly used as benchmark functions in multi-fidelity simulation studies. For example, one of them is the Borehole model \cite{bingham2024} which is an eight-dimensional problem and simulates water flow through a borehole. The analytical expressions and the coarse version of the benchmark examples are given in Appendix \ref{append:test_fun_fidelity}. The size of initial design to emulate $f$ and $\tilde f$ is $3d$ and $10d$, respectively. These points are generated based on LHS design in the same way as the single-fidelity case. A larger initial DoE is considered for $\tilde f$ given that it is computationally less demanding. We continue the sampling procedure until the total number of evaluations to $f$ and $\tilde f$ reaches $27d$. This makes our results consistent with the single-fidelity framework presented in Section \ref{sec:results}. 

\begin{figure}[htpb] 
	\includegraphics[width=0.49\textwidth]{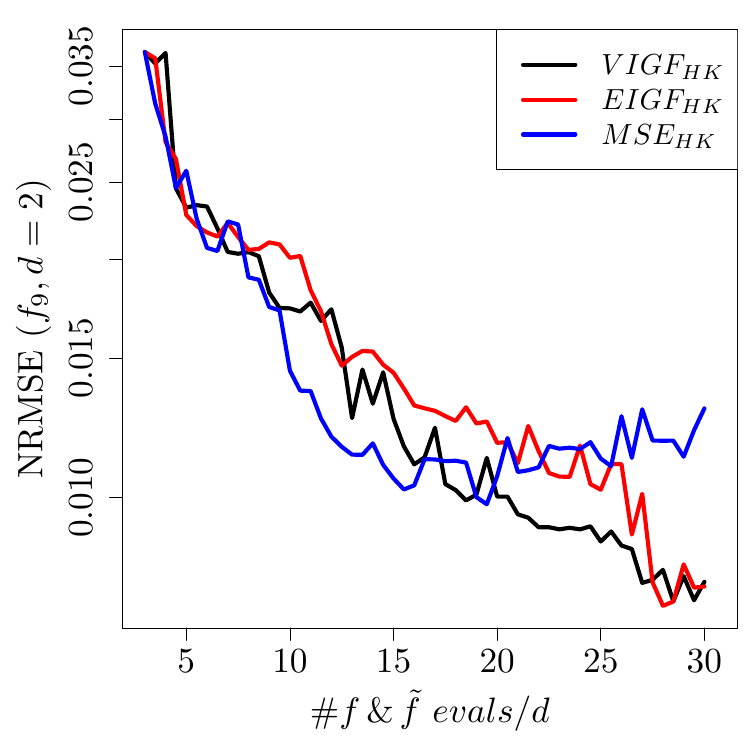} 
	\hspace{0.1cm}
	\includegraphics[width=0.49\textwidth]{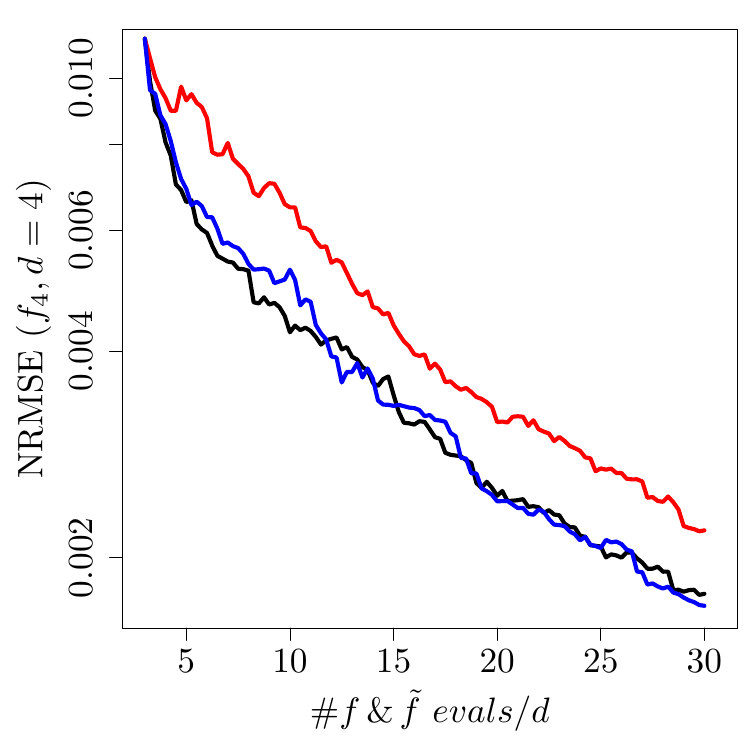} \\
	\includegraphics[width=0.49\textwidth]{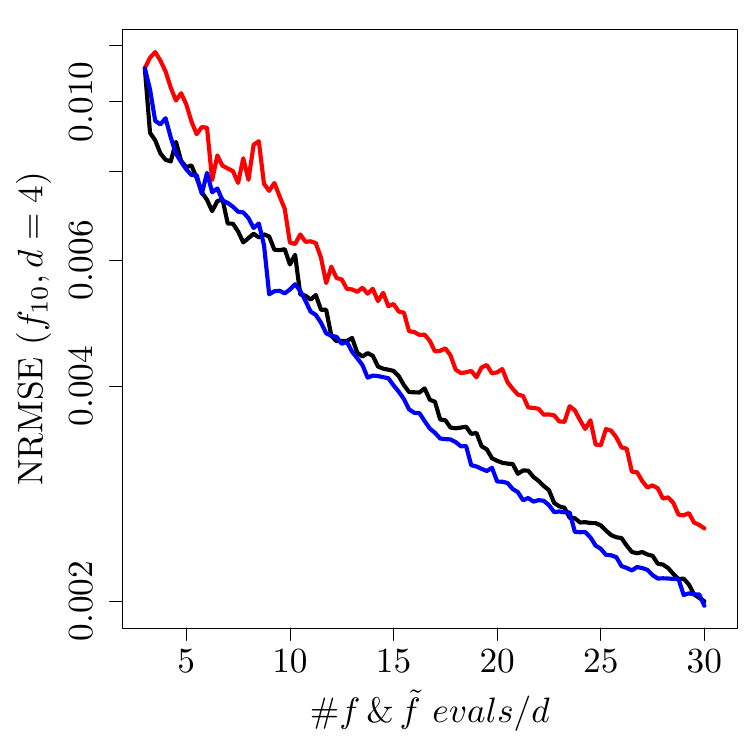}
	\hspace{0.1cm}
	\includegraphics[width=0.49\textwidth]{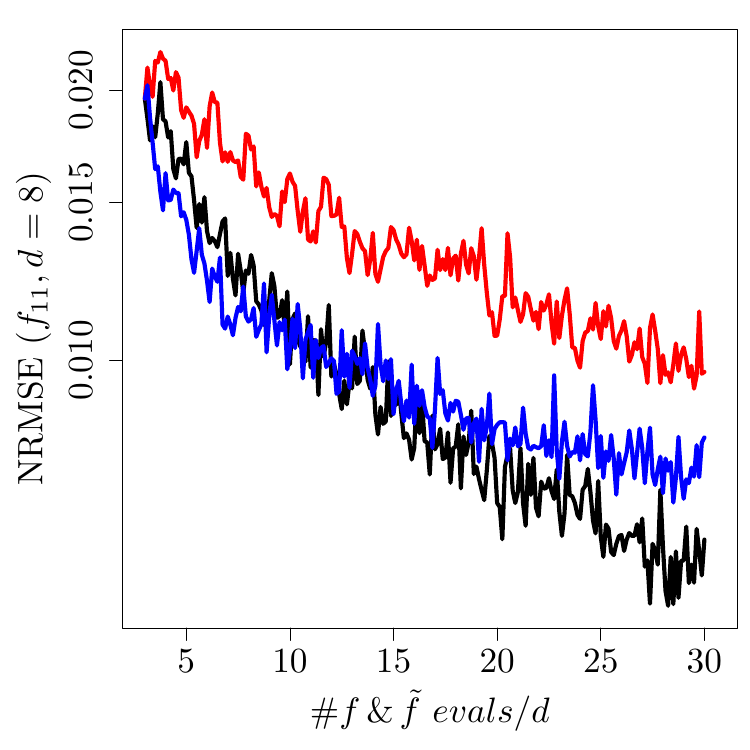} 
	\caption{Accuracy results associated with \text{$VIGF_{HK}$} (black), \text{$EIGF_{HK}$} (red), and \text{$MSE_{HK}$} (blue). Each curve is the median of ten NRMSEs. The $x$-axis represents the total number of evaluations to $f$ and $\tilde f$ divided by the problem dimension. The $y$-axis is on a logarithmic scale.}
	\label{res_HK}
\end{figure}

Figure \ref{res_HK} shows the performance of the \text{$VIGF_{HK}$} (black), \text{$EIGF_{HK}$} (red), and \text{$MSE_{HK}$} (blue) methods on the four test functions. Each curve represents the median of ten NRMSEs obtained using ten different initial DoE. Here, NRMSE is a measure of the distance between the emulator and high-fidelity function $f$. As can be seen, \text{$VIGF_{HK}$} is the winning algorithm in general; it has a superior performance over \text{$EIGF_{HK}$} in all cases. The accuracy obtained by \text{$VIGF_{HK}$} and \text{$MSE_{HK}$} is comparable on test functions $f_4$ and $f_{10}$. However, \text{$VIGF_{HK}$} works better on $f_9$ and $f_{11}$. 
\section{Conclusions}
\label{sec:conclusion}
This paper presents a simple, yet effective GP-based adaptive sampling algorithm for predicting the output of computationally expensive computer codes. The method selects a new sample and add it to the current design $\mathbf{X}_n$ sequentially to improve the emulator using VIGF. This criterion is the variance of an improvement function defined at any $\bx\in\mathcal{D}$ as the square of the difference between the fitted GP emulator and the model output at the nearest site in $\mathbf{X}_n$. The proposed selection criterion tends to fill the domain with a focus on ``interesting" regions, where the model output changes drastically. Such property is due to an efficient trade-off between exploration and exploitation. We suggested the batch mode of VIGF by multiplying it by the repulsion function relying on the GP correlation function. Moreover, we proposed the extension of VIGF (and also EIGF and MSE) to the  two-level fidelity case where in addition to the high-fidelity simulator, we have access to a coarse model which is cheap-to-evaluate. This is performed via the hierarchical kriging methodology. The performance of our method is evaluated on several commonly used benchmark functions and the accuracy results are compared with a few sampling strategies. We observed that our algorithm generally outperforms the other approaches in emulating the test cases. 

As a future research direction, the VIGF criterion could be extended to incorporate constraints, as explored in \cite{angun2023}. This will improve the applicability of VIGF to a wider range of adaptive design problems. Another important direction for future work is to extend the VIGF criterion to handle high-dimensional problems. However, as the number of design variables increases, the number of training points required to build a reliable emulator also grows. This issue becomes particularly challenging when working with costly computer models, where increasing the number of training data may not be feasible due to the high computational burden.
\begin{appendices}
\section{Test functions (single-fidelity)}
\label{append:test_fun}
\begin{enumerate}
\item $f_1(\bx) = 0.75\exp \left(-\frac{(9x_1 - 2)^2}{4} -\frac{(9x_2 - 2)^2}{4} \right) + 0.75\exp \left(-\frac{(9x_1 + 1)^2}{49}  -\frac{9x_2 + 1)}{10} \right)  
 \newline +  0.5\exp \left(-\frac{(9x_1 - 7)^2}{4} -\frac{(9x_2 - 3)^2}{4} \right) - 0.2 \exp \left( -(9x_1 - 4)^2 - (9x_2 - 7)^2 \right)$
 \item $f_2(\bx) = 4\left(x_1 - 2 + 8x_2 - 8x_2^2\right)^2 + \left(3 - 4x_2\right)^2 + 16\sqrt{x_3 + 1}\left(2x_3 - 1\right)^2 .$
 \item $f_3(\bx) = - \sum_{i=1}^{4} \boldsymbol{\alpha}_i \exp\left( \sum_{j= 1}^{3} \mathbf{A}_{ij} \left( x_j - \mathbf{P}_{ij} \right)^2 \right)$ where $\boldsymbol{\alpha} = (1, 1.2, 3, 3.2)^\top$, \\
 \[ \mathbf{A} = 
 \begin{bmatrix}
 3    & 10 & 30 \\
 0.1 & 10 & 35 \\
 3    & 10 & 30 \\
 0.1 & 10 & 35
 \end{bmatrix}
 , \quad \mathbf{P} = 10^{-4}
 \begin{bmatrix}
 3689 & 1170  & 2673\\
 4699 & 4387 & 7470 \\
 1091  & 8732 & 5547  \\
 381   &  5743 & 8828
 \end{bmatrix}
 \]
 \item $f_4 (\bx) = \frac{x_1}{2}\left( \sqrt{1 + (x_2 + x_3^2)\frac{x_4}{x_1^2}} - 1 \right) + (x_1+ 3x_4)\exp\left(1 + \sin(x_3) \right)$
 \item $f_5(\bx) = 10\sin(\pi x_1x_2) + 20(x_3 - 0.5)^2 +10x_4 + 5x_5$
 \item $f_6(\bx) = \exp \left( \sin \left( [0.9 (x_1 + 0.48))]^{10} \right)  \right) + x_2x_3 + x_4$
 \item $f_7(\bx) = \frac{(V_{b} + 0.74)x_6(x_5 + 9)}{x_6(x_5 + 9) + x_3} +  \frac{11.35x_3}{x_6(x_5 + 9) + x_3} + \frac{0.74 x_3x_6(x_5 + 9)}{(x_6(x_5 + 9) + x_3)x_4}$  where $V_{b} = \frac{12x_2}{x_1 + x_2}$ with:
\begin{itemize}
\item $x_1 \in \left[50, 150\right]$ is the resistance $b_1$ (K-Ohms)
\item $x_2 \in \left[25, 70\right]$ is the resistance $b_2$ (K-Ohm)
\item $x_3 \in \left[0.5, 3\right]$ is the resistance $f$ (K-Ohms)
\item $x_4 \in \left[1.2, 2.5\right]$ is the resistance $c_1$ (K-Ohms)
\item $x_5 \in \left[0.25, 1.2\right]$ is the resistance $c_2$ (K-Ohms)
\item $x_6 \in \left[50, 300\right]$ is the current gain $c_1$ (amperes)
\end{itemize}
\item $f_8(\bx) = 2\pi \sqrt{\frac{x_1}{x_4 + x_2^2 \frac{x_5 x_3 x_6}{x_7 V^2}}}$ where $V = \frac{x_2}{2x_4} \left( \sqrt{A^2 + 4x_4\frac{x_5 x_3}{x_7}x_6} - A\right)$ and \newline $A = x_5 x_2 + 19.62x_1 - \frac{x_4 x_3}{x_2}$ with: 
 \begin{itemize}
 	\item $x_1 \in \left[30, 60\right]$ is the piston weight (kg)
 	\item $x_2 \in \left[0.005, 0.020\right]$ is the piston surface area ($m^2$)
 	\item $x_3 \in \left[0.002, 0.010\right]$ is the initial gas volume ($m^3$)
 	\item $x_4 \in \left[1000, 5000 \right]$ is the spring coefficient ($N/m$)
 	\item $x_5 \in \left[90000, 110000 \right]$ is the atmospheric pressure ($N/m^2$)
 	\item $x_6 \in \left[290, 296\right]$ is the ambient temperature (K)
 	\item $x_7 \in \left[340, 360 \right]$ is the  filling gas temperature (K)
 \end{itemize}
\end{enumerate}
\section{Test functions (two-level fidelity)}
\label{append:test_fun_fidelity}
\begin{enumerate}
\item $\tilde f_4 (\bx) = \left(1 + \frac{\sin(x_1)}{10}\right)f_4 (\bx) - 2x_1 + x_2^2 + x_3^2 + 0.5$
\item $f_9(\bx) = \left(1 - \exp\left(-\frac{1}{2x_2}\right)\right) \frac{2300x_1^3 + 1900x_1^2 + 2092x_1 + 60}{100x_1^3 + 500x_1^2 + 4x_1 + 20}$
\item $\tilde f_9(\bx) = \frac{1}{4}\left[f_9(x_1+0.05, x_2+0.05) +  f_9(x_1+0.05, \max(0, x_2 - 0.05)) \right] 
\newline + \frac{1}{4}\left[f_9(x_1-0.05, x_2+0.05) +  f_9(x_1-0.05, \max(0, x_2 - 0.05)) \right]$
\item $f_{10}(\bx) = \frac{2}{3}\exp(x_1+x_2) - x_4\sin(x_3) + x_3$
\item $\tilde f_{10}(\bx) = 1.2f_{10}(\bx) - 1$
\item $f_{11}(\bx) = \frac{2\pi x_1 (x_2 - x_3)}{\ln(x_4/x_5)\left(1 + \frac{2x_6x_1}{\ln(x_4/x_5)x_5^2x_7} + \frac{x_1}{x_8}\right)}$ with:
\begin{itemize}
 	\item $x_1 \in \left[63070, 115600\right]$ is transmissivity of upper aquifer ($m^2/yr$)
	\item $x_2 \in \left[990, 1110\right]$ is potentiometric head of upper aquifer ($m$)
	\item $x_3 \in \left[700, 820\right]$ is potentiometric head of lower aquifer ($m$)
	\item $x_4 \in \left[100, 50000 \right]$ is radius of influence ($m$)
	\item $x_5 \in \left[0.05, 0.15\right]$ is radius of borehole ($m$)
	\item $x_6 \in \left[1120, 1680\right]$ is length of borehole ($m$)
	\item $x_7 \in \left[9855, 12045\right]$ is hydraulic conductivity of borehole ($m/yr$)
	\item $x_8 \in \left[63.1, 116\right]$ is transmissivity of lower aquifer ($m^2/yr$)
\end{itemize}
\item $\tilde f_{11}(\bx) = \frac{5 x_1 (x_2 - x_3)}{\ln(x_4/x_5)\left(1.5 + \frac{2x_6x_1}{\ln(x_4/x_5)x_5^2x_7} + \frac{x_1}{x_8}\right)}$
\end{enumerate}
\end{appendices}
\bibliography{biblio}
\bibliographystyle{plain}
\end{document}